\documentstyle[12pt,epsfig]{article}
\begin{document}
\pagestyle{empty}

\begin{center} 
{\Large {\bf Spontaneous breaking of a global symmetry in the 
Minimal Supersymmetric $SU(3)_{C}\times SU(3)_{L}\times U(1)_{N}$ Model.}}
\end{center}

\begin{center}
M. C. Rodriguez  \\
{\it Grupo de F{\'{\i}}sica Te\'{o}rica e Matem\'{a}tica F\'{\i}sica \\
Departamento de F\'{\i}sica  \\
Universidade Federal Rural do Rio de Janeiro - UFRRJ \\
BR 465 Km 7, 23890-000, Serop\'{e}dica - RJ \\
Brasil}
\end{center}

\date{\today} 

\begin{abstract}
We present a preliminar study of the scalar potential of the usual scalars in the  Minimal Supersymmetric $SU(3)_{C}\times SU(3)_{L}\times U(1)_{N}$ 
Model. We will consider the case in which all the usual neutral 
scalars fields of this model obtain vacuum expectation values. When we 
allow in our superpotential, only interactions that respect the invariance 
of the quantum number, ${\cal F}\equiv B+L$, where $B$ is the baryon 
number while $L$ is the total lepton number. We obtain, as in the minimal 
$SU(3)_{C}\times SU(3)_{L}\times U(1)_{N}$ model without supersymmetry, a 
Majoron. However, by allowing the violation  of this quantum number, we can 
remove the Majoron in a mechanism similar to that recently presented in the 
Supersymmetric version of the scheme of Gelmini-Roncadelli. The mass 
spectrum obtained is in agreement with current experimental data. 
\end{abstract}

PACS number(s): 12.60. Cn, 12.60. Jv


\section{Introduction}

In the Standard Model (SM) \cite{sg,Kronfeld:2010bx}, the relationship 
among the isospin $I_{3}$, the hypercharge $Y$ and the eletromagnetic 
charge $Q$ is defined by the Gell-Mann-Nishijima equation, which is 
defined as 
\begin{equation}
Q=I_{3}+Y.
\end{equation}
In the SM, we introduce only left-handed neutrinos\footnote{We do not to 
introduce right-handed neutrinos, 
$\left[ \nu_{iR}\sim \left( {\bf 1},{\bf 1},0 \right) \right]$, in 
agreement with the two component theory of neutrino theory of 
Landau \cite{Landau}, Lee and Yang \cite{LeeYang} and Salam \cite{Salam} 
from 1957, later the helicity of neutrinos was measured experimentally in 
1958 and it confirmed this hypothesis \cite{Goldhaber:1958nb}. 
Those right-handed neutrinos are known as sterile neutrinos 
\cite{Volkas:2001zb,Gonzalez-Garcia:2022pbf,Bilenky:2012qb,Petcov:2019yud}.} 
as follows \cite{quigg,donoghue,chengli}
\begin{eqnarray}
L_{iL}&=&\left( 
\begin{array}{c}
\nu_{iL} \\ 
l_{iL}         
\end{array} 
\right) \sim \left( {\bf 1},{\bf 2},- \frac{1}{2} \right), \,\ 
l_{iR}\sim \left( {\bf 1},{\bf 1},-1 \right), \,\ i=1,2,3.  
\label{lsm}
\end{eqnarray}
We show in parenthesis the transformations properties under 
the respective factors $(SU(3)_{C},SU(2)_{L},U(1)_{Y})$. The quark 
sector is defined as
\begin{eqnarray}
Q_{iL}&=&\left( 
\begin{array}{c}
u_{iL} \\ 
d_{iL}         
\end{array} 
\right) \sim \left( {\bf 3},{\bf 2},+ \frac{1}{6} \right), \,\ 
u_{iR}\sim \left( {\bf 3},{\bf 1},+ \frac{2}{3} \right), \,\
d_{iR}\sim \left( {\bf 3},{\bf 1},- \frac{1}{3} \right). \nonumber \\  
\label{qsm}
\end{eqnarray}  

In the SM, we have to introduce the following scalar field $\phi$ in the 
doublet representation of the $SU(2)_{L}$ \cite{Pich:1994zt,herrero}
\begin{eqnarray}
\phi &=&\left( 
\begin{array}{c}
\phi^{+} \\ 
\phi^{0}          
\end{array} 
\right)\sim \left( {\bf 1},{\bf 2},+ \frac{1}{2} \right), \,\ \Rightarrow \,\
\tilde{\phi} \equiv \left[ \left( i \sigma_{2} \right) \phi^{*} \right] = 
\left( 
\begin{array}{c}
\left( \phi^{0} \right)^{*} \\ 
- \phi^{-}         
\end{array} 
\right)\sim \left( {\bf 1},{\bf 2},- \frac{1}{2} \right). \nonumber \\
\label{hsm}
\end{eqnarray}
This scalar fields get the following vacuum expectation value (VEV)
\begin{eqnarray}
\langle \phi \rangle \equiv \frac{1}{\sqrt{2}}\left( 
\begin{array}{c}
0 \\ 
v          
\end{array} 
\right), \,\
\langle \tilde{\phi} \rangle = \frac{1}{\sqrt{2}}\left( 
\begin{array}{c}
v \\ 
0          
\end{array} 
\right).
\label{vevhiggsdomp}
\end{eqnarray}
The relevant terms in the SM lagrangian are given by
\begin{equation}
{\cal L}_{SM}={\cal L}^{Y}_{SM}- V(\phi) + {\cal L}^{kin}_{SM}. 
\end{equation}

The charged fermions obtain their masses, via the following Yukawa coupling \cite{quigg,donoghue,chengli}
\begin{eqnarray}
{\cal L}^{Y}_{SM}&=&\left[ g^{l}_{ij} 
\left( \bar{L}_{iL}\phi \right) l_{jR}+g^{d}_{ij} 
\left( \bar{Q}_{iL}\phi \right) d_{jR}+g^{u}_{ij}
\left( \bar{Q}_{iL}\tilde{\phi} \right) u_{jR}+hc \right].
\label{yukawaMP}
\end{eqnarray}

The scalar potential of the SM is defined in the following way 
\cite{Pich:1994zt,herrero}
\begin{equation}
V(\phi)= \mu^{2}|\phi|^{2}+ \lambda |\phi|^{4}.
\label{pothiggsmecanismo}
\end{equation}
Both the $\mu$-parameter and the $\lambda$-parameter are arbitrary and 
they are free parameters within the SM. For $\mu$-parameter, we can have 
the following two possibilities \cite{Pich:1994zt,herrero}
\begin{itemize}
\item $\mu^{2} >0:$ The gauge symmetry 
$SU(3)_{C}\times SU(2)_{L}\times U(1)_{Y}$ is not broken;
\item $\mu^{2} <0:$ We break the 
$SU(3)_{C}\times SU(2)_{L}\times U(1)_{Y}$ gauge symmetry to 
$SU(3)_{C}\times U(1)_{em}$. 
\end{itemize}
When we impose the condition of extremes
\begin{eqnarray}
\left. \frac{\partial V(\phi)}{\partial \phi}
\right|_{\langle \phi \rangle =v}=0, 
\end{eqnarray}
the minimum of the potential, in the case where $\mu^{2} <0$, is written 
in the following way
\begin{eqnarray}
v&\equiv& \sqrt{\frac{- \mu^{2}}{\lambda}}= \frac{1}{\sqrt{\sqrt{2}G_{F}}}=
246 \,\ GeV, \nonumber \\
G_{F}&=&\left( 1.16639 \pm 0.00002 \right) \times 10^{-5} \,\ GeV^{-2} 
\approx \frac{1}{(293 \,\ GeV)^{2}}.
\label{vevsm}
\end{eqnarray}
We make the following shift in the neutral scalar field
\begin{equation}
\phi^{0} \equiv \frac{1}{\sqrt{2}}\left( v+h^{0} + i G^{0} \right),
\label{hsm}
\end{equation}
under this expansion, the scalar potential of SM become \cite{Pich:1994zt}
\begin{eqnarray}
V(\phi)&=&V(\phi^{\prime 0})- \mu^{2}\left( h^{0} \right)^{2}+ \lambda vh^{0} 
\left( (h^{0})^{2}+ (G^{0})^{2} \right) + \frac{\lambda}{4} \left(
(h^{0})^{2}+ (G^{0})^{2} \right)^{2}, \nonumber \\ 
V(\phi^{\prime 0})&=&- \frac{\lambda}{4}v^{4}, \,\ 
|\phi^{\prime 0}|\equiv \frac{v}{\sqrt{2}}= \sqrt{\frac{- \mu^{2}}{2 \lambda}}>0.
\end{eqnarray}
The field $h^{0}$ is CP even field and gets the following mass
\footnote{A nice review about this interesting subject can be found in 
\cite{herrero}.}
\begin{equation}
M^{2}_{h^{0}}=-2 \mu^{2}=2 \lambda v^{2}.
\label{hmassSM}
\end{equation}
We must emphasize, the SM do not predict any value for this mass, in fact 
it can get any values from $0$ until a superior limit given by 
$M_{h^{0}}<1$ TeV \cite{Lee:1977eg,Lee:1977yc}. This mechanism is known 
as Brout-Englert-Higgs mechanism \cite{herrero,higgs}. 

The mass of the light Higgs boson $h^{0}$ was recently measured 
experimentally by ATLAS (A Toroidal LHC Apparatus) and 
CMS (Compact Muon Solenoid) collaborations, at Large Hadron Collider (LHC). 
Their measurements for the mass of the Higgs is \cite{pdg,Aad:2015zhl}
\begin{equation}
M_{h^{0}} = \left( 125.20  \pm 0.11 \right) \,\ GeV.
\label{expvalh}
\end{equation}

After breaking the gauge symmetry the gauge boson get mass, through the 
following term
\begin{eqnarray}
{\cal L}^{kin}_{SM}&=& 
\left( D_{m} \phi \right)^{\dagger}\left( D^{m} \phi \right), \nonumber \\
D_{m} \phi &=&\left[
\partial_{m}-ig 
\sum_{i=1}^{3}\left( \frac{\sigma^{i}}{2} \right) W^{i}_{m}- 
ig^{\prime}\left( \frac{1}{2} \right)W_{m} \right] \phi , 
\end{eqnarray}
where $W^{i}_{m}$ is the $SU(2)_{L}$ gauge bosons and $W_{m}$ 
is the $U(1)_{Y}$ gauge boson, while $g$ and $g^{\prime}$ are the gauge 
constant of the group $SU(2)_{L}$ and $U(1)_{Y}$, respectively. We obtain in addition to photon $A_{m}$, two massive bosons 
$W^{\pm}_{m}$ and $Z^{0}_{m}$ and their masses have the following 
expressions\footnote{The numerical value of the 
first relation below was obtained from \cite{average}} 
\cite{quigg,donoghue}
\begin{eqnarray}
M^{2}_{W^{\pm}}&=&\frac{g^{2}v^{2}}{4}=
\left[ \left( 80.3505 \pm 0.0077 \right) \,\ GeV \right]^{2},  
\nonumber \\
M^{2}_{Z^{0}}&=& \left( \frac{g^{2}+(g^{\prime})^{2}}{4} \right)v^{2}=
\frac{g^{2}v^{2}}{4}\left( 1+ \tan^{2}\theta_{W} \right)=
\frac{M^{2}_{W^{\pm}}}{\cos^{2}\theta_{W}}, \nonumber \\
W^{\pm}_{m}&=& \frac{1}{\sqrt{2}}\left( W^{1}_{m} \mp W^{1}_{m} \right), \,\
\left( 
\begin{array}{c}
Z^{0}_{m} \\ 
A_{m}          
\end{array} 
\right)=\left( 
\begin{array}{cc}
\cos \theta_{W}&- \sin \theta_{W} \\ 
\sin \theta_{W}& \cos \theta_{W}          
\end{array} 
\right)\left( 
\begin{array}{c}
W_{3m} \\ 
W_{m}          
\end{array} 
\right), \nonumber \\
\label{wsmval}
\end{eqnarray}
where $\theta_{W}$ is the Weinberg angle and it is defined as follows 
\begin{eqnarray}
\tan \theta_{W} \equiv \frac{g^{\prime}}{g}, \,\ \Rightarrow \,\ 
e=g \sin \theta_{W}=g^{\prime}\cos \theta_{W}.
\label{weinbergangledefinition}
\end{eqnarray}
The SM Lagrangian have four free parameters: $g,g^{\prime}, \mu^{2}$ and 
$\lambda$ in such a way, we can reproduce the old low-energy data \cite{quigg,donoghue,Pich:1994zt}.

The average of the experimental values for the gauge bosons masses 
are \cite{pdg,average}
\begin{eqnarray}
\left( M_{W^{\pm}} \right)_{ave}&=& \left(
80.4133 \pm 0.0080 \right) GeV, \nonumber \\
\left( M_{Z^{0}} \right)_{ave}&=& \left(
91.1875 \pm 0.0021 \right) GeV, 
\label{average}
\end{eqnarray}
while for the Weinberg angle, $\theta_{W}$, we have
\begin{eqnarray}
\sin^{2} \theta_{W}&=&0.2324 \pm 0.0012; \,\ \Rightarrow \,\ 
\cos^{2} \theta_{W}\equiv \left( \frac{M_{W^{\pm}}}{M_{Z^{0}}} \right)^{2} 
\sim 0.77.
\label{weibangexp}
\end{eqnarray}

Today we know neutrinos are massive particles 
\cite{Cribier:2019ckv,bilenkyprehist,bilenky-majorana,
bilenky-status,ion1,ion2}. 
If we introduce sterile neutrinos in SM, $\nu_{iR}$, we can write the 
following Dirac mass term for neutrinos 
\begin{equation}
{\cal L}^{\nu}_{\phi ,L}= y^{\nu}_{ij}\left[ 
\left( \bar{L}_{iL}\tilde{\phi} \right) \nu_{jR}+hc 
\right],
\label{neutrinodirac}
\end{equation} 
where $\tilde{\phi}$ is defined in our Eq.(\ref{hsm}). This mass 
term conserve lepton number defined in our Eq.(\ref{accidentalsyminSM}) and 
in this case the neutrinos are Dirac-type particles. A priori, to solve 
this problem, we can also to introduce the following Majorana mass term 
for the left-handed neutrinos
\cite{bilenky-majorana}
\begin{equation}
m^{M}_{ij}\left( \overline{\nu^{c}_{iL}}\nu_{jL}+hc \right).
\label{intgelmini}
\end{equation} 
This term break lepton number conservation, remember the SM  respects the following 
accidental global symmetry \cite{Gonzalez-Garcia:2022pbf}
\begin{equation}
G^{global}_{SM}\equiv U(1)_{B}\times U(1)_{L_{e}}\times U(1)_{L_{\mu}}
\times U(1)_{L_{\tau}},
\label{accidentalsyminSM}
\end{equation}
where $B$ is the baryon number and $L_{e},L_{\mu}$ and $L_{\tau}$ are 
the lepton number of each lepton.

An interesting way to generate the Eq.(\ref{intgelmini}), is to add the 
following term for the Yukawa coupling, defined by Eq.(\ref{yukawaMP}), 
lagrangian of SM \cite{chengli} 
\begin{equation}
\overline{L^{c}_{iL}}L_{jL}\sim \left( {\bf 1}, {\bf 1 \oplus 3},-1 \right); 
\,\ 
L^{c}_{iL}=C\bar{\psi}^{T}\sim \gamma_{2}f^{*}=\left( 
\begin{array}{c}
\nu^{c}_{iL} \\ 
l^{c}_{iL}         
\end{array} 
\right) \sim \left( {\bf 1}, {\bf 2},+ \frac{1}{2} \right).
\label{expgelminior}
\end{equation}
for $f= \nu$ and $e$. Therefore, to provide mass to neutrinos, we need to 
introduce new fields within the SM context. We can add to the SM singlets 
and also the triplets, of scalars and/or fermions, in representation of 
$SU(2)_{L}$. When we introduce
\begin{itemize}
\item Extra fermions in singlet, we generate masses for neutrinos by the known 
type I Seesaw mechanism \cite{seesaw1a,seesaw1b,seesaw1c};
\item Extra  scalars in triplets, it arise type II Seesaw mechanism 
\cite{seesaw2a,seesaw2b,seesaw2c,seesaw2d,seesaw2e}; 
\item Extra fermions in triplets, we obtain type III Seesaw mechanism \cite{seesaw3}.
\end{itemize} 
This new fields, will generate {\it Leptogenesis}
\cite{yanagida,Law:2009vh,Law:2010zz} an attractive scenario to explain the 
baryon asymmetry of the Universe \cite{Rodriguez:2016esw,Rodriguez:2020fvo}.

We can break the lepton number symmetry\footnote{The total lepton number 
$L$ is defined in terms of $L_{e},L_{\mu}$ and $L_{\tau}$, see 
Eq.(\ref{accidentalsyminSM}), in the following way $L=L_{e}+L_{\mu}+L_{\tau}$.} 
when we introduce the new scalar 
$\Delta$ in the triplet representation\footnote{The triplet $\Delta$ 
carries the lepton number of $2$.}, see Eq.(\ref{expgelminior}), of the 
group $SU(2)_{L}$
\begin{equation}
\Delta = \left( 
\begin{array}{cc}
\frac{h^{+}}{\sqrt{2}} & H^{++} \\
\Delta^{0} &- \frac{h^{+}}{\sqrt{2}}
\end{array} 
\right) \sim \left({\bf 1},{\bf3}, +1 \right), \,\ 
\langle \Delta \rangle = \frac{1}{\sqrt{2}} \left( 
\begin{array}{cc}
0 & 0 \\
V_{\Delta} & 0
\end{array} 
\right).
\label{tripletogelmini}
\end{equation}
This scalar allows us to write the following Yukawa interactions for neutrinos 
\cite{chengli,Gelmini:1980re}
\begin{eqnarray}
{\cal L}^{\nu}_{\Delta ,L}&=&g^{\nu}_{ij} \left[ 
\left( \overline{L^{c}_{iL}}\Delta L_{jL} \right) +hc \right] 
\nonumber \\
&=&g^{\nu}_{ij}\left[  \overline{\nu^{c}_{iL}}\nu_{jL}\Delta^{0}+ 
\left( 
\overline{\nu^{c}_{iL}}l_{jL}- \overline{l^{c}_{iL}}\nu_{jL}  
\right)\frac{h^{+}}{\sqrt{2}}+
\overline{l^{c}_{iL}}l_{jL}H^{++} +hc \right], \nonumber \\
\label{yukawatermogelmini}
\end{eqnarray}
the Yukawa coupling $h_{ij}$ is the $3 \times 3$ complex symmetric matrix.
Those scalars $h^{+}, \,\ H^{++}$ and $\Delta^{0}$ are known as bileptons 
\cite{Cuypers:1996ia}. The first term in the above equation is exactly our 
Eq.(\ref{intgelmini}), where we can identify
\begin{equation} 
m^{M}_{ij}=g^{\nu}_{ij} \frac{V_{\Delta}}{\sqrt{2}}.
\label{neutrinoGRscheme}
\end{equation}
It is know as Gelmini-Roncadelli scheme \cite{Gelmini:1980re}, it is 
type II Seesaw mechanism \cite{seesaw2a,seesaw2b,seesaw2c,seesaw2d,seesaw2e} and neutrinos are 
Majorana-type particles rather than being a Dirac-type particle such as all charged fermions. The double beta 
decay experiments, is defined in the following way
\begin{equation}
(A,Z) \rightarrow (A,Z+2)+2e^{-},
\label{doublebetadecaysemneut}
\end{equation}
can show if the neutrinos are Majorana particles \cite{Majorana} or 
Dirac particles \cite{Dirac}, see our Eq.(\ref{neutrinodirac}), more 
information on this subject can be seen at 
\cite{Bilenky:2012qb,Petcov:2019yud,bilenky-majorana,Barabash:2010bjn,Barabash:2011fg,Barabash:2023dwc}.

A useful parameter for placing strong limits in physics beyond the SM is 
the so-called $\rho$-parameter, this parameter is defined as follows
\begin{equation}
\rho \equiv 
\left( \frac{M^{2}_{W^{\pm}}}{M^{2}_{Z^{0}}\cos^{2} \theta_{W}} \right), 
\,\ \Rightarrow \rho^{SM}=1,
\label{rhoSM}
\end{equation} 
which is in perfect agreement with its 
experimental value given by \cite{pdg}
\begin{equation}
\rho_{\rm EXP} =0.9998\pm 0.0008.
\label{smexpbosonsector}
\end{equation}   

\begin{figure}[ht]
\begin{center}
\vglue -0.009cm
\mbox{\epsfig{file=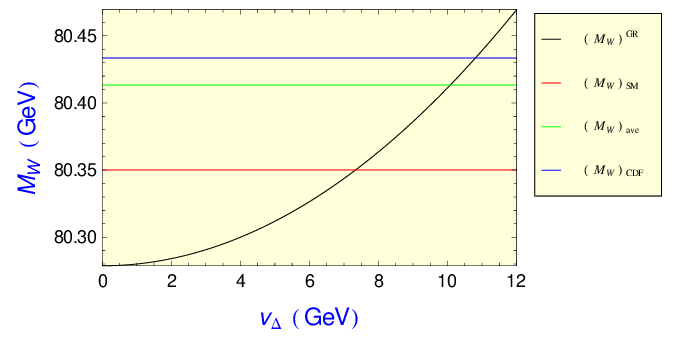,width=0.7\textwidth,angle=0}}       
\end{center}
\caption{We show the prediction about the behaviour for the gauge boson mass, 
$M_{W^{\pm}} \equiv M_{W}$, in the scheme of Gelmini-Roncadelli as a function of the 
Triplet VEV $V_{\Delta}$ (black curve) and we take $v=246$ GeV. The 
straight lines in red, green and blue are the values show in our  
Eqs.(\ref{wsmval},\ref{average},\ref{cdfresult}), respectively.}
\label{figmwgelmini}
\end{figure}

Using the new scalar triplet, $\Delta$ defined in our 
Eq.(\ref{tripletogelmini}), we can in addition to generating masses for 
neutrinos, jointly explain the recent measurement for $W$-boson mass, 
obtained by CDF collaboration \cite{cdf}
\begin{equation}
\left( M_{W^{\pm}} \right)_{CDF}= \left(
80.4335 \pm 0.0094 \right) \,\ GeV,
\label{cdfresult}
\end{equation}
because the masses of the gauge bosons in the scheme of Gelmini-Roncadelli  
are written as follows \cite{Montero:1999mc,carlosmajoron,Pisano:1997hk}
\begin{eqnarray}
M^{2}_{W^{\pm}}&=& \frac{g^{2}}{4}\left( v^{2}+2 V^{2}_{\Delta} \right)=
\frac{g^{2}v^{2}}{4}\left( 1+2R \right), \,\ 
R= \left( \frac{V_{\Delta}}{v} \right)^{2}, 
\nonumber \\
M^{2}_{Z^{0}}&=& \left( \frac{g^{2}+(g^{\prime})^{2}}{4} \right)
\left( v^{2}+4 V^{2}_{\Delta} \right)
= \frac{g^{2}v^{2}}{4\cos^{2}\theta_{W}}\left( 1+4R \right).
\label{gaugemassgelmini}
\end{eqnarray}
We can explain the new measurement for the 
$W$-boson mass if $V_{\Delta} \approx 10$ GeV, as can be easily seen in our 
Fig.(\ref{figmwgelmini}).

\begin{figure}[ht]
\begin{center}
\vglue -0.009cm
\mbox{\epsfig{file=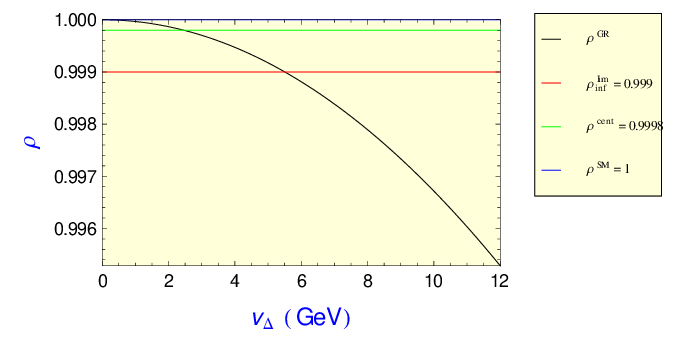,width=0.7\textwidth,angle=0}}       
\end{center}
\caption{We show the prediction about the behaviour for the $\rho$-parameter 
in the scheme of Gelmini-Roncadelli, see our Eq.(\ref{rhoparametergelmini}), 
as a function of the Triplet VEV $V_{\Delta}$ (black curve) and we take 
$v=246$ GeV. The straight lines in red, green and blue are the following 
values $\rho^{inf}= \rho^{cen}-0.0008$, $\rho^{cen}= 0.9998$ and 
$\rho^{SM}=1$ (as defined in our Eq.(\ref{rhoSM})), respectively.}
\label{figrhogelmini}
\end{figure}

On the other hand, the $\rho$-parameter in the scheme of 
Gelmini-Roncadelli, see Eq.(\ref{neutrinoGRscheme}), becomes 
\cite{Rodriguez:2022hsj}
\begin{eqnarray}
\rho &=& \frac{1+2R}{1+4R} \approx 1-2R.
\label{rhoparametergelmini}
\end{eqnarray}
With this expression, it is clear that we can explain the current experimental value of 
the $\rho$-parameter when $V_{\Delta} \approx 4.92$ GeV, see our 
Fig.(\ref{figrhogelmini}).  

In the scheme of Gelmini-Roncadelli, the scalar potential is written in 
the following form \cite{Ma:1998dx,Frampton:2002rn}
\begin{eqnarray}
V_{G}( \phi , \Delta)&=& \mu^{2}( \phi^{\dagger}\phi )+ 
M^{2}Tr( \Delta^{\dagger}\Delta )+ 
\lambda_{1}( \phi^{\dagger}\phi )^{2}+ 
\lambda_{2} \left[ Tr( \Delta^{\dagger}\Delta ) \right]^{2} 
\nonumber \\ &+&
\lambda_{3}Tr[( \Delta^{\dagger}\Delta )( \Delta^{\dagger}\Delta )] +
\lambda_{4}( \phi^{\dagger}\phi ) Tr( \Delta^{\dagger}\Delta) 
\nonumber \\ &+&
i \lambda_{5} \epsilon_{ijk} ( \phi^{\dagger} \sigma^{i}\phi ) 
(\Delta^{j \dagger}\Delta^{k}) . 
\label{potmajorongelmini}
\end{eqnarray}
Now, if we make the shift defined in our Eq.(\ref{hsm}) together with 
the following expansion
\begin{eqnarray}
\Delta^{0}=\frac{1}{\sqrt{2}} \left( 
V_{\Delta}+ R_{\Delta}+ i I_{\Delta} \right).
\label{expescgalminirocandelli}
\end{eqnarray} 

\begin{figure}[ht]
\begin{center}
\vglue -0.009cm
\mbox{\epsfig{file=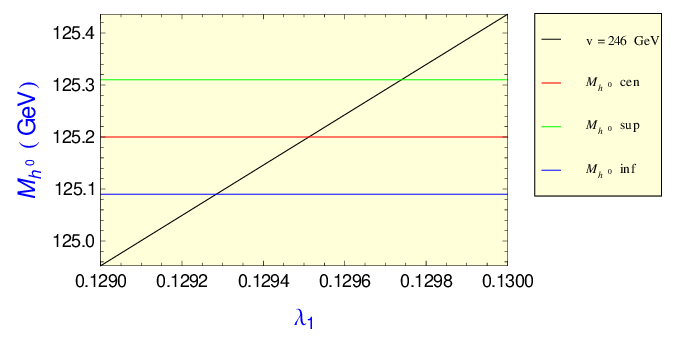,width=0.7\textwidth,angle=0}}       
\end{center}
\caption{We show the prediction about the behaviour for the lightest Higgs mass, 
$M_{h^{0}}$, in the scheme of Gelmini-Roncadelli, see our Eq.(\ref{hgelmini}), 
as function of the $\lambda_{1}$-parameter (black curve) and $v=246$ GeV. 
The  $M^{exp}_{h^{0}}$ is the central experimental values (red line), given at 
our Eq.(\ref{expvalh}), and $M^{sup}_{h^{0}}=M^{exp}_{h^{0}}+0.11$ (green line) 
while $M^{inf}_{h^{0}}=M^{exp}_{h^{0}}-0.11$ (blue line).}
\label{figmhgelmini}
\end{figure}

We get the following constraint equations
\begin{eqnarray}
\frac{t_{H}}{v}&=& \left( \mu^{2}+ \lambda_{1} v^{2}+ 
\frac{\lambda_{3}}{2}v^{2}_{\Delta}- 
\frac{\lambda_{4}}{2}v^{2}_{\Delta} \right)=0, \nonumber \\
\frac{t_{\Delta}}{V_{\Delta}}&=& \left( M^{2}+ \lambda_{2} v^{2}_{\Delta}+ 
\frac{\lambda_{3}}{2}v^{2}_{\Delta}- 
\frac{\lambda_{4}}{2}v^{2}_{\Delta}+
\lambda_{5}v^{2}_{\Delta} \right)=0.
\end{eqnarray}
After carrying out the shift, defined in our 
Eqs.(\ref{hsm},\ref{expescgalminirocandelli}), we obtain the following 
mass matrices for the neutral scalars in the scheme of Gelmini-Roncadelli
\begin{eqnarray}
\frac{1}{2} \left(
\begin{array}{cc}
R_{H} & R_{\Delta}
\end{array} 
\right) M^{2}_{R} \left(
\begin{array}{c}
R_{H} \\
R_{\Delta}
\end{array} 
\right) 
+ \frac{1}{2} \left(
\begin{array}{cc}
I_{H} & I_{\Delta}
\end{array} 
\right) M^{2}_{I} \left(
\begin{array}{c}
I_{H} \\
I_{\Delta}
\end{array} 
\right). 
\end{eqnarray}
For the scalar of CP even, we get the following mass matrix
\begin{equation}
M^{2}_{R}= \left(
\begin{array}{cc}
2 \lambda_{1} v^{2} & 
(\lambda_{3}- \lambda_{4})vV_{\Delta} \\
(\lambda_{3}- \lambda_{4})vV_{\Delta} & 
2 \lambda_{2}V^{2}_{\Delta}
\end{array} 
\right).
\label{cppargelmini}
\end{equation}
Where we get the following analytical results 
$det \left( M^{2}_{R} \right) \neq 0$ and 
$Tr \left( M^{2}_{R} \right) \neq 0$, therefore we have two massive 
scalars. The lightest Higgs, we will represent as $h^{0}$, and we also 
have another heavier Higgs, which we will denote as $H^{0}$. The mass of 
the lightest scalar, is the same as the Higgs boson of SM, compare with 
Eq.(\ref{hmassSM}), is\footnote{The complete expression for the mass of 
these scalars can be found in \cite{Yagyu:2012qp}. The mass of heavier 
Higgs, $H^{0}$, we will present below.}
\begin{equation}
M^{2}_{h^{0}} \approx 2 \lambda_{1}v^{2}.
\label{hgelmini}
\end{equation}
The mass $M_{h^{0}}$ as function of the $\lambda_{1}$-parameter is shown in 
our Fig.(\ref{figmhgelmini}). Where, we can see that we can get the central 
data of the lightest Higgs mass if we assume $\lambda_{1} \approx 0.1295$.

For the case of CP odd sector, we obtain the following mass matrix
\begin{equation}
M^{2}_{I}= \left(
\begin{array}{cc}
\frac{t_{H}}{v} & 0 \\
0 & \frac{t_{\Delta}}{V_{\Delta}}
\end{array} 
\right).
\end{equation}
We get two Goldstone bosons, because $det \left( M^{2}_{I} \right) = 0$ and 
$Tr \left( M^{2}_{I} \right) = 0$, one of them is the same $G^{0}$ obtained 
in SM. The extra Goldstone boson is the Majoron, 
$M^{0}\equiv I_{\Delta}$ \cite{Gelmini:1980re}. The Majoron, 
$M^{0}$, can be detected experimentally through the following process 
\cite{Barabash:2010bjn,Barabash:2011fg,Barabash:2023dwc}
\begin{equation}
(A,Z) \rightarrow (A,Z+2)+2e^{-}+M^{0}.
\label{majorondetection}
\end{equation}
The existence of a non-massive Majoron, $M^{0}$, is excluded by experimental 
data\footnote{Decay $Z^{0} \rightarrow M^{0}+X$.} \cite{nir}. One possible solution 
for this problem, is to make the Majoron, $M^{0}$, invisible to the leptons and 
quarks, but we will not use this solution in this article.

\begin{figure}[ht]
\begin{center}
\vglue -0.009cm
\mbox{\epsfig{file=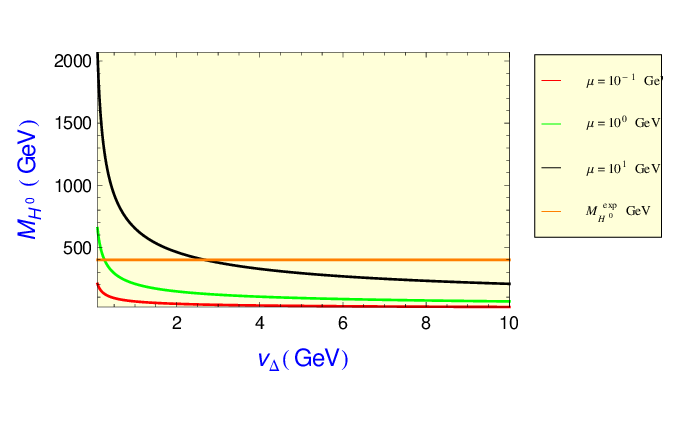,width=0.7\textwidth,angle=0}}       
\end{center}
\caption{We show the prediction about the behaviour for the heavier Higgs, 
$M_{H^{0}}$, in the scheme of Ma, see our Eq.(\ref{Hframpton}), as a function 
of the Triplet VEV $V_{\Delta}$, and $v=246$ GeV. The values of 
$\mu \equiv \mu_{M}$ show in the box in several colors. $M^{exp}_{H^{0}}$ is the 
lower experimental values given at our Eq.(\ref{limATLASH}) in orange line.}
\label{figmHgelmini}
\end{figure}

There are another three possible solutions to remove the non-massive 
Majoron \cite{Frampton:2002rn}
\begin{itemize}
\item By Higgs mechanism, $R_{\Delta}$ and $I_{\Delta}$ have to be absorbed;
\item Giving a large mass to the Majoron ($M^{0}$) without VEV;
\item Giving a large mass to the Majoron ($M^{0}$) with tiny VEV.
\end{itemize}
Let's quickly review the type of the last solution, where we need to add the 
following new term in the scalar potential
\footnote{We will denote it as scheme of Ma.} \cite{Ma:1998dx,Frampton:2002rn}
\begin{eqnarray}
V( \phi , \Delta)&=&V_{G}( \phi , \Delta) + \left( 
\mu_{M}\left( \phi^{\dagger} \Delta \tilde{\phi} \right)+ hc \right)  .
\label{potmajoronmp}
\end{eqnarray}
Where $V_{G}( \phi , \Delta)$ is defined in our Eq.(\ref{potmajorongelmini}) 
and $\tilde{\phi}$ is defined in our Eq.(\ref{hsm}). $\langle \Delta \rangle$ 
give mass only for neutrinos, review our Eq.(\ref{neutrinoGRscheme}), then it 
has tiny VEV, which is precisely the third case mentioned above.  In the scheme of Ma, 
our Eq.(\ref{cppargelmini}) become 
\begin{equation}
M^{2}_{R}= \left(
\begin{array}{cc}
2 \lambda_{1} v^{2} & 
(\lambda_{3}- \lambda_{4})vV_{\Delta}+2 \sqrt{2}\mu_{M}v \\
(\lambda_{3}- \lambda_{4})vV_{\Delta}+2 \sqrt{2}\mu_{M}v & 
2 \lambda_{2}V^{2}_{\Delta}- \frac{\mu_{M}}{\sqrt{2}} \frac{v^{2}}{V_{\Delta}}
\end{array} 
\right).
\label{cppargelmini}
\end{equation}
The mass spectrum, we get beyond the lightest Higgs, $h^{0}$, given in 
Eq.(\ref{hgelmini}), we obtain the heavier Higgs field, $H^{0}$, and its mass 
is given by
\begin{equation}
M^{2}_{H^{0}} \approx \frac{\mu_{M}}{\sqrt{2}} \frac{v^{2}}{V_{\Delta}}.
\label{Hframpton}
\end{equation} 
$M_{H^{0}}$ as function of, the VEV, $V_{\Delta}$ is shown in our 
Fig.(\ref{figmHgelmini}) and we obtain masses within experimental 
limits for the heavy Higgs \cite{ATLAS:2024jja}
\begin{eqnarray}
400 \,\ GeV \,\ \leq M_{H^{0}} \leq 1000 \,\ GeV, 
\label{limATLASH}
\end{eqnarray}
if $V_{\Delta} \leq 3$ GeV and $\mu \geq 10^{0}$ GeV.

\begin{figure}[ht]
\begin{center}
\vglue -0.009cm
\mbox{\epsfig{file=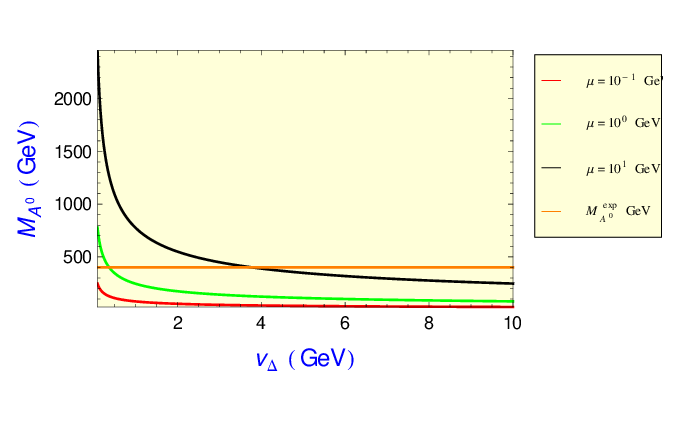,width=0.7\textwidth,angle=0}}       
\end{center}
\caption{We show the prediction about the behaviour for the pseudoscalar, 
$M_{A^{0}}$, in the scheme of Ma, see our Eq.(\ref{Mframpton}), as a function 
of the Triplet VEV $V_{\Delta}$, and $v=246$ GeV. The values of 
$\mu \equiv \mu_{M}$ show in the box in several colors. $M^{exp}_{A^{0}}$ is the lower 
experimental values given at our Eq.(\ref{limATLASA}) in orange line.}
\label{figmMma}
\end{figure}

The mass matrix for neutral CP odd, we get one massive pseudoscalar, $A^{0}$, 
\cite{Ma:1998dx,Frampton:2002rn}
\begin{equation} 
M^{2}_{A^{0}} \approx \frac{\mu_{M}}{\sqrt{2}} 
\left( \frac{4v^{2}_{\Delta}+v^{2}}{V_{\Delta}} \right)
=M^{2}_{H^{0}}+2 \sqrt{2}\mu_{M}V_{\Delta},
\label{Mframpton}
\end{equation}
therefore $M_{A^{0}}>M_{H^{0}}$. We show $M_{A^{0}}$ as function of, the 
VEV, $V_{\Delta}$ in our Fig.(\ref{figmMma}) and again our results 
for the mass of $M_{A^{0}}$ is in agreement experimental limits 
for the pseudoscalar, $A^{0}$, given by the following constraint 
\cite{ATLAS:2024jja}
\begin{eqnarray}
400 \,\ GeV \,\ \leq M_{A^{0}} \leq 1000 \,\ GeV, 
\label{limATLASA}
\end{eqnarray}
if $V_{\Delta} \leq 3$ GeV and $\mu \geq 10^{0}$ GeV.

\begin{figure}[ht]
\begin{center}
\vglue -0.009cm
\mbox{\epsfig{file=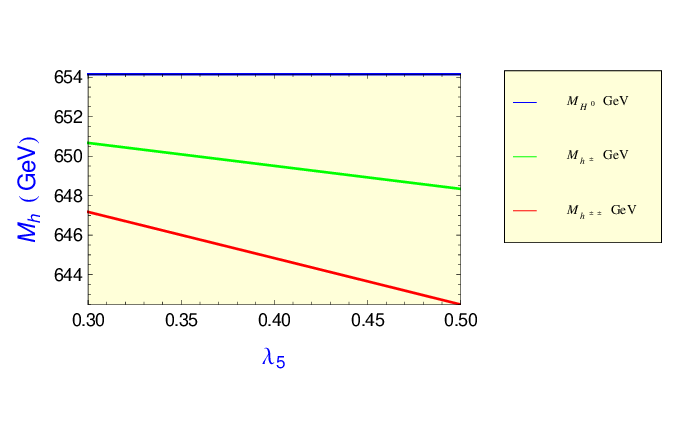,width=0.7\textwidth,angle=0}}       
\end{center}
\caption{We show the prediction about the behaviour for the masses,  
$M_{H^{0}}$ (blue curve), and $v=246$ GeV. The values of 
$\mu \equiv \mu_{M}$ show in the box in several colors. The $M_{h^{\pm}}$ (green curve) 
and $M_{h^{\pm \pm}}$ (red curve) when $\lambda_{5}>0$.}
\label{figmhl5p}
\end{figure}

\begin{figure}[ht]
\begin{center}
\vglue -0.009cm
\mbox{\epsfig{file=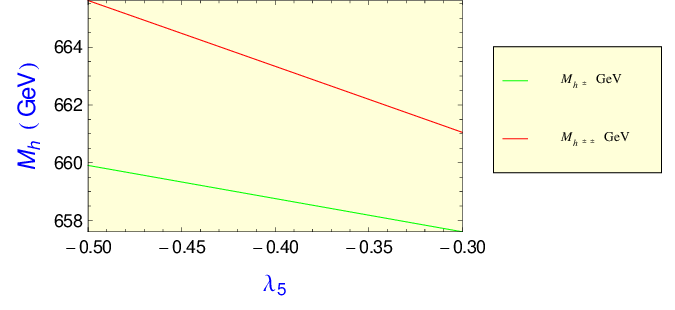,width=0.7\textwidth,angle=0}}       
\end{center}
\caption{We show the prediction about the behaviour for the masses 
$M_{h^{\pm}}$ (green curve) and $M_{h^{\pm \pm}}$ (red curve) when 
$\lambda_{5}<0$ and $M_{H^{0}}=654.150$ GeV. The values contained in this 
graph are in agreement with the analytical results obtained by 
Eqs.(\ref{mhl5p},\ref{numvall5n}).}
\label{figmhl5n}
\end{figure}

In the scheme of Ma, see our Eq.(\ref{potmajoronmp}), we also have charged 
scalars and their masses are given by the following relations 
\cite{Yagyu:2013kva,Staub:2014pca}
\begin{eqnarray}
M^{2}_{h^{\pm}}= M^{2}_{H^{0}}- \frac{\lambda_{5}}{4}v^{2}, \,\
M^{2}_{h^{\pm \pm}}= M^{2}_{H^{0}}- \frac{\lambda_{5}}{2}v^{2}.
\label{gelminicarregados}
\end{eqnarray} 
We have the following hierarchy if $\lambda_{5}>0$
\begin{equation}
M_{H^{0}}>M_{h^{\pm}}>M_{h^{\pm \pm}}, \,\ \mbox{see our Fig.(\ref{figmhl5p})},
\label{mhl5p}
\end{equation}
while for the case of $\lambda_{5}<0$ we have
\begin{equation}
M_{h^{\pm \pm}}>M_{h^{\pm}}>M_{H^{0}}, \,\ \mbox{see our Fig.(\ref{figmhl5n})}.
\label{mhl5n}
\end{equation}
Our numerical results satisfy the following 
experimental constraints $M_{h^{\pm}}>155$ GeV \cite{pdg} and 
$M_{h^{\pm \pm}}>445$ GeV \cite{Bambhaniya:2014cia}. We want to 
emphasize that the masses of the charged scalars for the case
$\lambda_{5}<0$ are heavier than in the opposite case.

Supersymmetry, it is more known as SUSY, arose in theoretical papers more 
than 30 years ago independently by Golfand and Likhtman \cite{gl},
Volkov and Akulov \cite{va} and Wess and Zumino \cite{wz}.
The  supersymmetry algebra was introduced in \cite{gl} \footnote{They also 
constructed the first four-dimensional field theory with supersymmetry, 
(massive) quantum electrodynamics (SQED) of spinors and scalars. We will call 
it as {\bf Golfand-Likhtman model} as suggested by Shiffman in his 
article \cite{Shifman:2015tka}.}. The 
Volkov-Akulov article, however, deals only with fermions and its name is 
``{\bf Is the Neutrino a Goldstone Particle?}"
\footnote{It is the base of supergravity.}. 
For interested people interested in the history of supersymmetry see 
\cite{volkov1,Likhtman:2001st} and for the earliest Supersymmetric Standard 
Model see \cite{fay1,marcoshist}, while anyone interested, how to work with 
supersymmetry see the great review article \cite{ogievetski}.

The Supersymmetric Standard Model (MSSM) is a good 
candidate to be the physics beyond the SM. In this 
model we can solve the hierarchy problem as well to explain the Higgs 
Masses\footnote{The current status of the search for supersymmetry is presented in reference \cite{gladkaza}.} 
\cite{kazakov1,Rodriguez:2019mwf}. In the MSSM, as in the SM, neutrinos are 
massless. If we want to give mass to the neutrinos, we must break $R$-Parity. 
In this mechanism, the masses generated are a combination of 
type-I \cite{seesaw1a,seesaw1b,seesaw1c} 
and type-III Seesaw \cite{seesaw3} mechanism. Another interesting result that 
arises when we break $R$-Parity is to explain the new $W$ mass 
get from the CDF collaboration \cite{Rodriguez:2021qec}.

There is also, the Minimal $R$-Symmetric extension of the Minimal 
Superymmetric Standard Model, known as MRSSM 
\cite{Kribs:2007ac,Diessner:2014ksa,Diessner:2019ebm}. The field content 
of the MRSSM is enlarge compared to the MSSM as we shown in our 
Tab.(\ref{table:mattermrssm})and the triplet $\hat{T}$ is defined as 
\cite{Diessner:2014ksa,Diessner:2019ebm}
\begin{eqnarray}
\hat{T}&=&\left( 
\begin{array}{cc}
\frac{\hat{T^{0}}}{\sqrt{2}} & \hat{T}^{+} \\ 
\hat{T}^{-} & \frac{\hat{T^{0}}}{\sqrt{2}}           
\end{array} 
\right).
\end{eqnarray} 

\begin{table}
\begin{tabular}{c|cc}
SuperfieldField & $(SU(3)_{C},SU(2)_{L})_{U(1)_{Y}}$  & $U(1)_{R}$ \\ \hline
$\hat{Q}_L$& $({\bf 3},{\bf 2})_{1/6}$ & 1 \\ 
$\hat{U}_R$& $({\bf \bar 3},{\bf 1})_{-2/3}$ & 1 \\ 
$\hat{D}_R$& $({\bf \bar 3},{\bf 1})_{1/3}$ & 1 \\ 
$\hat{L}_L$& $({\bf 1},{\bf 2})_{-1/2}$ & 1 \\ 
$\hat{E}_R$& $({\bf 1},{\bf 1})_{1}$ & 1 \\ 
$\hat{S}$ & $({\bf 1},{\bf 1})_{0}$ & 0 \\ 
$\hat{T}$ & $({\bf 1},{\bf 3})_{0}$& 0 \\ 
$\hat{{\cal O}}$ & $({\bf 8},{\bf 1})_{0}$& 0 \\ 
$\hat{H}_{1}$& $({\bf 1},{\bf 2})_{1/2}$ & 0 \\ 
$\hat{H}_{2}$& $({\bf 1},{\bf 2})_{-1/2}$ & 0 \\
$\hat{T}$& $({\bf 1},{\bf 3})_{0}$ & 0 \\
$\hat{S}$& $({\bf 1},{\bf 1})_{0}$ & 0 \\ 
$\hat{R}_{1}$& $({\bf 1},{\bf 2})_{-1/2}$ & 2 \\ 
$\hat{R}_{2}$& $({\bf 1},{\bf 2})_{+1/2}$ & 2  
\end{tabular}
\caption{Matter and $R$-charges in the $R$-symmetric supersymmetric
model, MRSSM.}
\label{table:mattermrssm}
\end{table}

\begin{figure}[ht]
\begin{center}
\vglue -0.009cm
\mbox{\epsfig{file=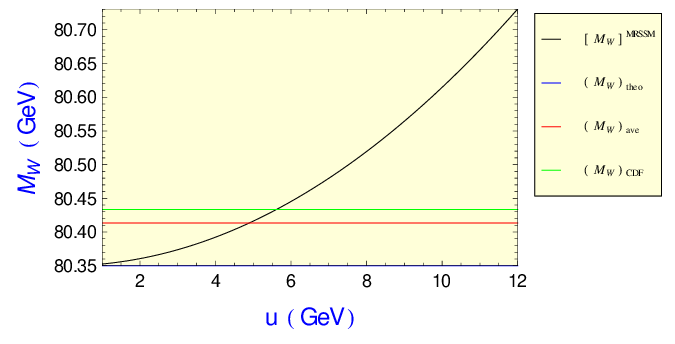,width=0.7\textwidth,angle=0}}       
\end{center}
\caption{We show the prediction about the behaviour for the gauge boson 
mass $W^{\pm}$ in MRSSM, see our Eq.(\ref{gaugemassMRSSM}), as a function 
of the Triplet VEV $v_{T}$. The straight lines in blue, red and green are 
the values show in our  Eqs.(\ref{wsmval},\ref{average},\ref{cdfresult}), 
respectively.}
\label{figmwmrssm}
\end{figure}

\begin{figure}[ht]
\begin{center}
\vglue -0.009cm
\mbox{\epsfig{file=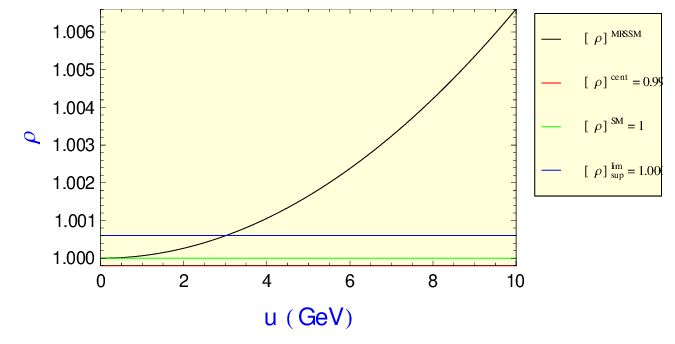,width=0.7\textwidth,angle=0}}       
\end{center}
\caption{We show the prediction about the behaviour for the $\rho$-parameter 
in MRSSM, see our Eq.(\ref{rhoMRSSM}), as a function of the Triplet VEV 
$v_{T}$. Where we use the following values 
$\rho^{cen}= 0.9998$, $\rho^{lim}_{sup}= \rho^{cen}+0.0008$ and $\rho^{SM}=1$ as 
defined in our Eq.(\ref{rhoSM}).}
\label{figrhomrssm}
\end{figure}

The superpotential of this model is \cite{Staub:2014pca}
\begin{equation}
W= \left( \mu_{1}+ \lambda_{1}\hat{S} \right) \hat{H}_{1}\hat{R}_{1}+
\left( \mu_{2}+ \lambda_{2}\hat{S} \right) \hat{H}_{2}\hat{R}_{2}+
\Lambda_{1}\hat{R}_{1}\hat{T}\hat{H}_{1}+
\Lambda_{2}\hat{R}_{2}\hat{T}\hat{H}_{2}.
\end{equation}
We can generate Majorana mass for neutrinos through the operator
\begin{equation}
H_{2}H_{2}L_{iL}L_{jL}.
\end{equation}
The mass of $W$ is given by \cite{Diessner:2014ksa,Diessner:2019ebm}
\begin{equation}
M^{2}_{W}= \frac{g^{2}}{4}\left( v^{2}+4v^{2}_{T} \right)=
\frac{g^{2}v^{2}}{4}\left( 1+R \right), \,\ 
R= \left( \frac{2v_{T}}{v} \right)^{2},
\label{gaugemassMRSSM}
\end{equation}
We can explain the new measurement for the 
$W$-boson mass, given at Eq.(\ref{cdfresult}), if $|v_{T}|\leq 4$ GeV \cite{Diessner:2014ksa,Diessner:2019ebm} and see our 
Fig.(\ref{figmwmrssm}). The mass expression for $Z$ gauge boson has 
the same expression as presented in our Eq.(\ref{wsmval}). The 
$\rho$-parameter is \cite{Diessner:2014ksa,Diessner:2019ebm}
\begin{equation}
\rho = \left( 1+ R \right), \,\ 
R= \left( \frac{2v_{T}}{v} \right)^{2},
\label{rhoMRSSM}
\end{equation} 
and see our Eq.(\ref{figrhomrssm}). The mass spectrum in Higgs sector 
was presented 
\cite{Diessner:2014ksa,Staub:2014pca,Diessner:2014ksa,Diessner:2015bna}.

There are also the supersymmetric version of the scheme of Gelmini-Roncadelli
\footnote{We will call this model as SUSYGR by short.} 
\cite{Rossi:2002zb,Brignole:2003iv,DAmbrosio:2004rko,Rossi:2005ue,Joaquim:2008bm}, 
where as in the MSSM, we have two Higgs in the doublet representations
\footnote{We need also to introduce their superpartners, known as higgsinos. 
They are put in the chiral superfields \cite{ogievetski,kazakov1,Rodriguez:2019mwf}.} 
\begin{eqnarray}
H_{1}&=&\left( 
\begin{array}{c}
h^{0}_{1} \\ 
h^{-}_{1}          
\end{array} 
\right)\sim \left( {\bf 1},{\bf 2},- \frac{1}{2} \right), \,\ 
\langle H_{1} \rangle = \frac{1}{\sqrt{2}}\left( 
\begin{array}{c}
v_{1} \\
0  
\end{array} 
\right), \nonumber \\
H_{2}&=&\left( 
\begin{array}{c}
h^{+}_{2} \\ 
h^{0}_{2}          
\end{array} 
\right)\sim \left( {\bf 1},{\bf 2},+ \frac{1}{2} \right), \,\ 
\langle H_{2} \rangle = \frac{1}{\sqrt{2}}\left( 
\begin{array}{c}
0 \\ 
v_{2}          
\end{array} 
\right).
\end{eqnarray}
We also add one scalar in the  triplet representation 
${\bf T}_{1}=(T_{1},T_{2},T_{3})$ and another one in the 
anti-triplet representation 
${\bf T}_{2}=( \bar{T}_{1}, \bar{T}_{2}, \bar{T}_{3})$, where
\begin{eqnarray}
\Delta_{1}&=& \left( 
\begin{array}{cc}
\Delta^{0}_{1} &- \frac{h^{+}_{1}}{\sqrt{2}} \\
- \frac{h^{+}_{1}}{\sqrt{2}} &- H^{++}_{1}
\end{array} 
\right) \sim \left({\bf 1},{\bf 3},+ 1 \right), \,\ 
\langle \Delta_{1} \rangle = \frac{1}{\sqrt{2}} \left( 
\begin{array}{cc}
V_{\Delta_{1}} & 0 \\
0 & 0
\end{array} 
\right), \nonumber \\
\Delta_{2}&=&\left( 
\begin{array}{cc}
\Delta^{--}_{2} &- \frac{h^{-}_{2}}{\sqrt{2}} \\
- \frac{h^{-}_{2}}{\sqrt{2}} &- \Delta^{0}_{2}
\end{array} 
\right) \sim \left({\bf 1},{\bf \bar{3}},- 1 \right), \,\ 
\langle \Delta_{2} \rangle = \frac{1}{\sqrt{2}} \left( 
\begin{array}{cc}
0 & 0 \\
0 & V_{\Delta_{2}}
\end{array} 
\right). \nonumber \\
\label{tripletosusygelmini}
\end{eqnarray}
The superpotential of the SUSYGR is written as follows 
\cite{Rossi:2002zb,Brignole:2003iv,DAmbrosio:2004rko,Rossi:2005ue,Joaquim:2008bm}
\begin{eqnarray}
W^{GR}_{SUSY}&=&W^{MSSM}_{RC}+
\mu_{\Delta}\left( \hat{\Delta}_{1}\hat{\Delta}_{2} \right) +
f^{N}_{ij}Tr \left[ \hat{L}_{i} \hat{\Delta}_{1} \hat{L}_{j} \right]+
\lambda_{1}Tr \left[ \hat{H}_{1} \hat{\Delta}_{1} \hat{H}_{1} \right] 
\nonumber \\ &+&
\lambda_{2}Tr \left[ \hat{H}_{2} \hat{\Delta}_{2}\hat{H}_{2} \right], \nonumber \\
W^{MSSM}_{RC}&=& \mu \left( \hat{H}_{1}\hat{H}_{2} \right) + 
f^{l}_{ij} \left( \hat{H}_{1}\hat{L}_{i} \right) \hat{E}_{j}+ 
f^{d}_{ij} \left( \hat{H}_{1}\hat{Q}_{i} \right) \hat{D}_{j}+
f^{u}_{ij} \left( \hat{H}_{2}\hat{Q}_{i} \right) \hat{U}_{j}. \nonumber \\
\label{suppotSUSYGR}
\end{eqnarray}
Where $\left( \hat{\Delta}_{1}\hat{\Delta}_{2} \right) \equiv 
\epsilon_{\alpha \beta}\hat{\Delta}_{1}^{\alpha}\hat{\Delta}_{2}^{\beta}$ and 
$W^{MSSM}_{RC}$ is the term already well known from MSSM with $R$ parity 
conservation and you can see this term in \cite{Rodriguez:2019mwf}. Therefore 
the mass mechanism for the neutrinos is the same as in the scheme of 
Gelmini-Roncadelli and it is presented in \cite{Rossi:2002zb,Rossi:2005ue}. 

The scalar field $\Delta_{2}$ is not necessary to give mass for any fermion in 
the SUSYGR. The terms proportional to $\lambda_{1}$ and $\lambda_{2}$ 
explicitly break lepton number conservation if we assign $L=-2 \,\ (2)$ to the 
triplet $\Delta_{1} \,\ ( \Delta_{2})$ \cite{Rossi:2002zb}. Therefore, 
these two parameters break the lepton number conservation, $L$, and they will 
allow us to implement the Scheme of Ma, see our Eq.(\ref{potmajoronmp}), in the 
SUSYGR.

\begin{figure}[ht]
\begin{center}
\vglue -0.009cm
\mbox{\epsfig{file=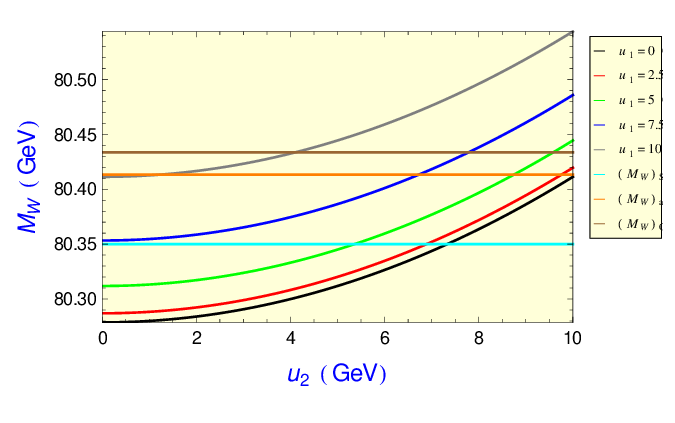,width=0.7\textwidth,angle=0}}       
\end{center}
\caption{We show the prediction about the behaviour for the gauge boson mass $W^{\pm}$ 
in SUSYGR, see our Eq.(\ref{gaugemassSusygelmini}), as a function of the Triplet VEV 
$u_{2} \equiv V_{\Delta_{2}}$, and some values of $u_{1} \equiv V_{\Delta_{1}}$ show 
in the box in several colors and $v=246$ GeV. The straight lines in brown, orange and cyan are the values show in our  Eqs.(\ref{wsmval},\ref{average},\ref{cdfresult}), 
respectively.}
\label{figmwsusygelmini}
\end{figure}

\begin{figure}[ht]
\begin{center}
\vglue -0.009cm
\mbox{\epsfig{file=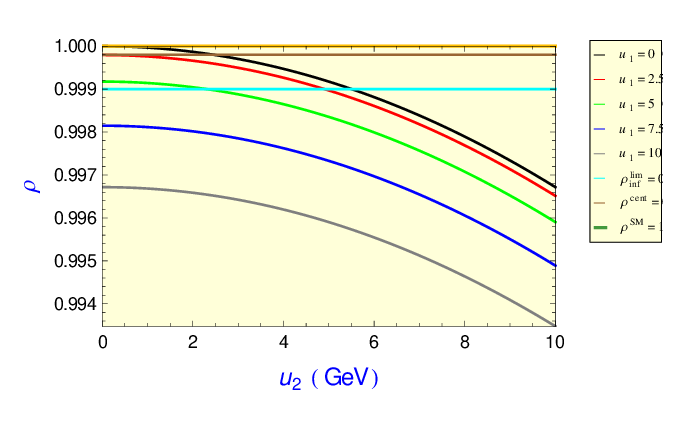,width=0.7\textwidth,angle=0}}       
\end{center}
\caption{We show the prediction about the behaviour for the $\rho$-parameter 
in SUSYGR, see our Eq.(\ref{gaugemassSusygelmini}), as a function of the Triplet VEV 
$u_{2} \equiv V_{\Delta_{2}}$, and some values of $u_{1} \equiv V_{\Delta_{1}}$ show 
in the box in several colors and $v=246$ GeV. Where we use the following values 
$\rho^{cen}= 0.9998$, $\rho^{inf}= \rho^{cen}-0.0008$ and $\rho^{SM}=1$ as 
defined in our Eq.(\ref{rhoSM}).}
\label{figrhosusygelmini}
\end{figure}

The masses of the gauge bosons in the SUSYGR can be written in the following 
way 
\begin{eqnarray}
M^{2}_{W^{\pm}}&=& \frac{g^{2}}{4}
\left[ v^{2}+2 \left( V^{2}_{\Delta_{1}}+ V^{2}_{\Delta_{2}} \right) \right]
= \frac{g^{2}v^{2}}{4}\left[ 1+2 \left( R_{1}+R_{2} \right) \right], 
\nonumber \\
M^{2}_{Z^{0}}&=& \frac{g^{2}}{4 \cos^{2}\theta_{W}} \left[ v^{2}+
4 \left( V^{2}_{\Delta_{1}}+ V^{2}_{\Delta_{2}} \right) \right]=
\frac{g^{2}v^{2}}{4 \cos^{2}\theta_{W}}\left[ 1+
4 \left( R_{1}+R_{2} \right) \right], \nonumber \\
\rho&=& \frac{1+2 \left( R_{1}+R_{2} \right)}{1+4 \left( R_{1}+R_{2} \right)}
\approx 1-2 \left( R_{1}+R_{2} \right), \nonumber \\
R_{1}&=& \left( \frac{V_{\Delta_{1}}}{v} \right)^{2}, \,\ 
R_{2}= \left( \frac{V_{\Delta_{2}}}{v} \right)^{2}.
\label{gaugemassSusygelmini}
\end{eqnarray}
Our numerical results are show in 
Figs.(\ref{figmwsusygelmini},\ref{figrhosusygelmini}). 
The usual scalar sector of this model is presented 
in \cite{Yagyu:2012qp,Yagyu:2013kva,Staub:2014pca,Brignole:2003iv}. The resolution of 
these two problems at the same time in this model, we believe, should be better studied 
later in this SUSYGR. 

Models with the following gauge symmetry 
\begin{equation}
SU(3)_{C}\times SU(3)_{L}\times U(1)_{N}
\end{equation} 
are known as $331$ for short. They are interesting  
possibilities for the physics at the TeV scale due the following nice 
aspects \cite{singer,ppf,331rh}
\begin{enumerate}
\item The family number must be a multiple of three in order to cancel 
anomalies \cite{ppf};
\item Why $\sin^{2} \theta_{W}<\frac{1}{4}$ is observed \cite{ppf};
\item It is the simplest model that includes bileptons of both types: 
scalars and vectors ones, see our Eq.(\ref{lq});
\item The models have a scalar sector similar to the two Higgs doublets 
Model, see Sec.(\ref{msusy331}) or our 
Eqs.(\ref{decindoublets},\ref{decindoublets1},\ref{decindoublets2},\ref{22HM}), 
which allow to predict the quantization of the electric charge \cite{pr};
\item It solves the strong $CP$-problem \cite{pal1};
\item The model has several sources of CP violation 
\cite{Montero:1998nf,Montero:1998yw,Montero:2005yb}.
\end{enumerate}

We will concentrate in the model presented in Ref. \cite{ppf}, and 
the solution to the anomaly for the 
mass of the $W$-boson in the $331$ was recently presented in the 
following reference \cite{VanLoi:2022eir}. The scalar sector of the 
minimal $331$ model, $m331$, was presented in 
\cite{Montero:1999mc,ppf,Pleitez:1993gc,tonasse}. We show the most general 
scalar potential in our Eq.(\ref{potentialm331}). For example the terms 
proportional to $(\eta S \eta^{\dagger}\eta)$ and $(\chi \rho SS)$ does not 
conserve lepton number conservation, and they are very similar to 
$\mu$, implemented the scheme of Ma, presented in our Eq.(\ref{potmajoronmp}). 

A few years ago, it was made a preliminary study about the Majoron in the 
$m331$, in this article they do not used interactions that break lepton number 
conservation \cite{Montero:1999mc}. Therefore, they do not remove the Majoron, 
$M^{0}$. However in the $m331$ the Majoron, $M^{0}$, does not couple, at the 
tree level, neither to charged leptons nor to the quarks, then we avoid 
the so dangerous decay $Z^{0} \rightarrow R_{\Delta}+M^{0}$ 
\cite{Montero:1999mc,deSPires:2003wwk}.

The minimal supersymmetric $SU(3)_{C}\times SU(3)_{L}\times U(1)_{N}$ model, 
or it is more known as MSUSY331 for shortness, was built in  
\cite{ema1,pal2,331susy1,mcr,Rodriguez:2005jt}. An interesting aspect of 
the model, is that the mass of the lightest Higgs is around $125.5$ GeV 
\cite{331susy1,Rodriguez:2005jt}. In this preliminary analysis, the 
hypothesis was used that only the scalars 
$\sigma^{0}_{2}$ and $\sigma^{\prime 0}_{2}$ get VEV. Now we will allow all the 
usual scalars of the MSUSY331 to have VEV, it means we want to consider 
that both $\sigma^{0}_{1}$ and $\sigma^{\prime 0}_{1}$ get VEV in similar 
conditions as presented in \cite{Rodriguez:2022hsj}. 

The goal of this work is to show that using a mechanism similar to scheme of 
Ma, see our Eq.(\ref{potmajoronmp}), in MSUSY 331, we were able to remove the 
Majoron from the model in a similar mechanism as presented in SUSYGR. We get 
this nice solution, because the neutral fields belong to the sextet,  
($\sigma^{\prime 0}_{1}$, $\sigma^{\prime 0}_{2}$), and to the
anti-sextet, ($\sigma^{0}_{1}$), representation of the group $SU(3)_{L}$, 
are not necessary to generate mass for any charged fermion in the model, 
so their vacuum expectation values can be a few GeV, but the pseudoscalar 
will get large mass around ${{\cal O}}(400)$ GeV and in this way removing the 
non-massive Majoron from MSUSY331.

\section{Minimal Supersymmetric $331$ Model.}
\label{msusy331}

We will introduce the following chiral superfields associated with leptons and quarks
$\hat{L}_{1,2,3}$, $\hat{Q}_{1,2,3}$, $\hat{u}^{c}_{1,2,3}$, 
$\hat{d}^{c}_{1,2,3}$, $\hat{J}^{c}$ and $\hat{j}^{c}_{1,2}$ \cite{331susy1,mcr}. 
The particle content of each chiral superfield and anti-chiral supermultiplet 
is presented in the Tabs.(\ref{lfermionnmssm},\ref{rfermionnmssm}), 
respectively. 

\begin{table}[h]
\begin{center}
\begin{tabular}{|c|c|c|}
\hline 
$\mbox{ Chiral Superfield} $ & $\mbox{ Fermion} $ & $\mbox{ Scalar} $ \\
\hline
$\hat{L}_{iL}=( \hat{\nu}_{i}, \hat{l}_{i}, \hat{l}^{c}_{i})^{T}_{L}
\sim({\bf 1},{\bf3},0)$ & 
$L_{iL}=(\nu_{i},l_{i},l^{c}_{i})^{T}_{L}$ & 
$\tilde{L}_{iL}=( \tilde{\nu}_{i}, \tilde{l}_{i}, \tilde{l}^{c}_{i})^{T}_{L}$ \\
\hline
$\hat{Q}_{1L}=(\hat{u}_{1}, \hat{d}_{1}, \hat{J})^{T}_{L}\sim
({\bf 3},{\bf3},+(2/3))$ & 
$Q_{1L}=(u_{1},d_{1},J)^{T}_{L}$ & 
$\tilde{Q}_{1L}=(\tilde{u}_{1}, \tilde{d}_{1}, \tilde{J})^{T}_{L}$ \\ 
\hline
\end{tabular}
\end{center}
\caption{\small Particle content in the chiral superfields in MSUSY331 
and we neglected the color indices and $i=1,2,3$. In parenthesis it appears 
the transformations properties under the respective factors 
$(SU(3)_{C},SU(3)_{L},U(1)_{N})$.}
\label{lfermionnmssm}
\end{table}

\begin{table}[h]
\begin{center}
\begin{tabular}{|c|c|c|}
\hline 
$\mbox{ Anti-Chiral Superfield} $ & $\mbox{ Fermion} $ & $\mbox{ Scalar} $ \\
\hline
$\hat{Q}_{\alpha L}=(\hat{d}_{\alpha}, \hat{u}_{\alpha}, \hat{j}_{m})^{T}_{L}
\sim({\bf 3},{\bf\bar{3}},-(1/3))$ & 
$Q_{\alpha L}=(d_{\alpha},u_{\alpha},j_{m})^{T}_{L}$ & 
$\tilde{Q}_{\alpha L}=
(\tilde{d}_{\alpha}, \tilde{u}_{\alpha}, \tilde{j}_{m})^{T}_{L}$ \\
\hline
$\hat{u}^{c}_{iL}\sim({\bf \bar{3}},{\bf1},-(2/3))$ & $u^{c}_{iL}\equiv \bar{u}_{iR}$ & 
$\tilde{u}^{c}_{iL}$ \\ 
\hline
$\hat{d}^{c}_{iL}\sim({\bf \bar{3}},{\bf1},+(1/3))$ & 
$d^{c}_{iL}\equiv \bar{d}_{iR}$ & 
$\tilde{d}^{c}_{iL}$ \\ 
\hline
$\hat{J}^{c}_{L}\sim({\bf \bar{3}},{\bf1},-(5/3))$ & 
$J^{c}_{L}\equiv \bar{J}_{R}$ & 
$\tilde{J}^{c}_{L}$   \\
\hline
$\hat{j}^{c}_{\beta L}\sim({\bf \bar{3}},{\bf1},+(4/3))$ & 
$j^{c}_{\beta L}\equiv \bar{j}_{\beta R}$ & 
$\tilde{j}^{c}_{\beta L}$   \\
\hline
\end{tabular}
\end{center}
\caption{\small Particle content in the anti-chiral superfields in 
MSUSY331 and $\alpha =2,3$, $i=1,2,3$ and $\beta =1,2$.}
\label{rfermionnmssm}
\end{table}

The usual scalars fields in the minimal, $m331$, model are defined as 
three triplets ($\eta, \rho, \chi$) and one anti-sextet ($S$) \cite{ppf} and 
we can decompose those fields in 
\begin{equation}
\left( SU(3)_{C} \times SU(3)_{L}\times U(1)_{N} \right) \supset 
\left( SU(3)_{C} \times SU(2)_{L}\times U(1)_{Y} \right)
\end{equation} 
representations in the following way \cite{Montero:1999mc,Liu:1993gy,Liu:1993fwa,DeConto:2015eia}
\begin{itemize}
\item 
$\left( {\bf 1}, {\bf 3},0 \right) \rightarrow 
\left( {\bf 1}, {\bf 2}, -1 \right) \oplus 
\left( {\bf 1}, {\bf 1}, +2 \right)$
\begin{eqnarray}
\eta &=& \left( 
\begin{array}{c} 
\Phi_{\eta} \\ 
\eta^{+}_{2}          
\end{array} \right) 
\sim ({\bf1},{\bf3},0), \nonumber \\
\Phi_{\eta} &=& \left( 
\begin{array}{c} 
\eta^{0} \\ 
\eta^{-}_{1}          
\end{array} \right) 
\sim ({\bf1},{\bf2},-1),\quad 
\eta^{+}_{2} 
\sim ({\bf 1},{\bf 1},+2), 
\label{decindoublets}
\end{eqnarray}
where the singlet $\eta^{+}_{2}$ was proposed by Zee \cite{Zee:1980ai}
\footnote{See the Eq.(\ref{expgelminior}).}
\item 
$\left( {\bf 1}, {\bf 3},+1 \right) \rightarrow 
\left( {\bf 1}, {\bf 2}, +1 \right) \oplus 
\left( {\bf 1}, {\bf 1}, +4 \right)$
\begin{eqnarray}
\rho &=& \left( 
\begin{array}{c} 
\Phi_{\rho} \\ 
\rho^{++}          
\end{array} \right) 
\sim ({\bf 1},{\bf 3},+1), \nonumber \\
\Phi_{\rho} &=& \left( 
\begin{array}{c} 
\rho^{+} \\ 
\rho^{0}          
\end{array} \right) 
\sim ({\bf 1},{\bf 2},+1), \quad
\rho^{++} \sim ({\bf 1},{\bf 1},+4),
\label{decindoublets1}
\end{eqnarray}
\item 
$\left( {\bf 1}, {\bf 3},-1 \right) \rightarrow 
\left( {\bf 1}, {\bf 2}, -3 \right) \oplus 
\left( {\bf 1}, {\bf 1}, 0 \right)$
\begin{eqnarray}
\chi &=& \left( 
\begin{array}{c} 
\Phi_{\chi} \\ 
\chi^{0}          
\end{array} \right) 
\sim ({\bf 1},{\bf 3},-1), \nonumber \\ 
\Phi_{\chi} &=& \left( 
\begin{array}{c} 
\chi^{-} \\ 
\chi^{--}          
\end{array} \right) 
\sim ({\bf 1},{\bf 2},-3), \quad
\chi^{0}\sim ({\bf 1},{\bf 1},0),
\label{decindoublets2}
\end{eqnarray}
\item $\left( {\bf 1}, {\bf \bar{6}},0 \right) \rightarrow 
\left( {\bf 1}, {\bf 3}, +2 \right) \oplus
\left( {\bf 1}, {\bf 2}, +1 \right) \oplus 
\left( {\bf 1}, {\bf 1}, +4 \right)$
\begin{eqnarray}
S&=& \left( 
\begin{array}{cc} 
T & \frac{\Phi_{S}}{\sqrt{2}} \\ 
\frac{\Phi^{T}_{S}}{\sqrt{2}} & H^{--}_{2}          
\end{array} \right)
\sim ({\bf 1},{\bf {\bf \bar{6}}},0), 
\nonumber \\
\Phi_{S}&=& \left( 
\begin{array}{c} 
h^{+}_{2} \\ 
\sigma^{0}_{2}          
\end{array} \right)
\sim ({\bf 1},{\bf 2},+1), \quad 
H^{--}_{2}
\sim ({\bf 1},{\bf 1},+4), 
\label{22HM}
\end{eqnarray}
\end{itemize}
where $\Phi_{S}$ is the Higgs doublet boson of SM, see Eq.(\ref{hsm}), and 
$H^{--}_{2}$ appear in the model of Babu presented in reference 
\cite{babuh2mm}. The triplet
\begin{eqnarray}
T&=&\left( \begin{array}{cc} 
\sigma^{0}_{1} & \frac{h^{+}_{1}}{\sqrt{2}} \\ 
\frac{h^{+}_{1}}{\sqrt{2}} & H^{--}_{1} \\
\end{array} \right)
\sim ({\bf 1},{\bf 3},+2),
\label{tripin6^{*}}
\end{eqnarray}
is the triplet $\Delta_{1}$, see Eq.(\ref{tripletosusygelmini})
\footnote{As we expect the MSUSY331 contain the SUSYGR, and the neutral scalar 
$\sigma^{0}_{1}$ as the scalar $\Delta^{0}_{1}$ generate Majorana mass term for 
the neutrinos \cite{Rodriguez:2022hsj}.}. We redefine the 
neutral components of the scalars defined above in the following way 
\begin{eqnarray} 
\eta^{0}&=&\frac{1}{\sqrt{2}}
\left( v_{\eta}+ R^{0}_{\eta}+i I^{0}_{\eta}  \right), \quad 
\rho^{0}= \frac{1}{\sqrt{2}}
\left( v_{\rho}+ R^{0}_{\rho}+i I^{0}_{\rho}  \right), \nonumber \\
\chi^{0}&=&\frac{1}{\sqrt{2}}
\left( v_{\chi}+ R^{0}_{\chi}+i I^{0}_{\chi} \right), \quad
\sigma^{0}_{1}=\frac{1}{\sqrt{2}}
\left( v_{\sigma^{0}_{1}}+ R^{0}_{\sigma^{0}_{1}}+i I^{0}_{\sigma^{0}_{1}}  
\right), \nonumber \\                   
\sigma^{0}_{2}&=&\frac{1}{\sqrt{2}}
\left( v_{\sigma^{0}_{2}}+ 
R^{0}_{\sigma^{0}_{2}}+i I^{0}_{\sigma^{0}_{2}} \right).
\label{vevantisextet} 
\end{eqnarray}

In order to implement supersymmetry, and also at same time to cancel chiral 
anomalies, we must to introduce the following scalars fields \cite{331susy1,mcr}
\begin{eqnarray}
\eta^{\prime} &=& 
\left( 
\begin{array}{c} 
\eta^{\prime 0} \\ 
\eta^{\prime +}_{1} \\
\eta^{\prime -}_{2}          
\end{array} \right) 
\sim ({\bf1},{\bf \bar{3}},0),\quad
\rho^{\prime} = 
\left( \begin{array}{c} 
\rho^{\prime -} \\ 
\rho^{\prime 0} \\
\rho^{\prime --}          
\end{array} 
\right) 
\sim ({\bf1},{\bf \bar{3}},-1), \nonumber \\ 
\chi^{\prime} &=& 
\left( \begin{array}{c} 
\chi^{\prime +} \\ 
\chi^{\prime ++} \\
\chi^{\prime 0}          
\end{array} \right) 
\sim ({\bf1},{\bf \bar{3}},+1), \quad
S^{\prime} = 
      \left( \begin{array}{ccc} 
\sigma^{\prime 0}_{1}& \frac{h^{\prime -}_{2}}{ \sqrt{2}}& 
\frac{h^{\prime +}_{1}}{ \sqrt{2}} \\ 
\frac{h^{\prime -}_{2}}{ \sqrt{2}}& H^{\prime --}_{1}& 
\frac{ \sigma^{ \prime 0}_{2}}{ \sqrt{2}} \\
\frac{h^{\prime+}_{1}}{ \sqrt{2}}& 
\frac{ \sigma^{ \prime 0}_{2}}{ \sqrt{2}}&  
H^{\prime ++}_{2}        
\end{array} \right) \sim ({\bf1},{\bf6},0), \nonumber \\
\label{3t} 
\end{eqnarray}
again we redefine
\begin{eqnarray} 
\eta^{\prime 0}&=&\frac{1}{\sqrt{2}}
\left( v_{\eta^{\prime}}+ 
R^{0}_{\eta^{\prime}}+i I^{0}_{\eta^{\prime}} \right), \quad 
\rho^{\prime 0}= \frac{1}{\sqrt{2}}
\left( v_{\rho^{\prime}}+ 
R^{0}_{\rho^{\prime}}+i I^{0}_{\rho^{\prime}} \right), \nonumber \\
\chi^{\prime 0}&=&\frac{1}{\sqrt{2}}
\left( v_{\chi^{\prime}}+ 
R^{0}_{\chi^{\prime}}+i I^{0}_{\chi^{\prime}} \right), \quad
\sigma^{\prime 0}_{1}=\frac{1}{\sqrt{2}}
\left( v_{\sigma^{\prime 0}_{1}}+ 
R^{0}_{\sigma^{\prime 0}_{1}}+i I^{0}_{\sigma^{\prime 0}_{1}} \right), \nonumber \\                   
\sigma^{\prime 0}_{2}&=&\frac{1}{\sqrt{2}}
\left( v_{\sigma^{\prime 0}_{2}}+ 
R^{0}_{\sigma^{\prime 0}_{2}}+i I^{0}_{\sigma^{\prime 0}_{2}} \right).
\label{vevantisextet} 
\end{eqnarray}

In the MSUSY331 we need to introduce the following three vector superfields 
$\hat{V}^{a}_{C}\sim({\bf 8},{\bf 1}, 0)$
\footnote{The gluinos are the superpartner of gluons, and therefore they are in the adjoint representation of $SU(3)$, which is real.}, 
where $a=1,2, \ldots ,8$, $\hat{V}^{a}\sim({\bf 1},{\bf 8}, 0)$, and 
$\hat{V}\sim({\bf 1},{\bf 1}, 0)$. The particle content 
in each vector superfield is presented in the Tab.(\ref{gaugemssm}). 
The Lagrangian of this model is presented in the references 
\cite{331susy1,mcr}.
\begin{table}[h]
\begin{center}
\begin{tabular}{|c|c|c|c|}
\hline 
${\rm{Vector \,\ Superfield}}$ & ${\rm{Gauge \,\ Bosons}}$ & ${\rm{Gaugino}}$ & ${\rm Gauge \,\ constant}$ \\
\hline 
$\hat{V}^{a}_{C}\sim({\bf 8},{\bf 1}, 0)$ & $g^{a}_{m}$ & $\lambda _{C}^{a}$ & $g_{s}$ \\
\hline 
$\hat{V}^{a}\sim({\bf 1},{\bf 8}, 0)$ & $V^{a}_{m}$ & $\lambda^{a}$ & $g$ \\
\hline
$\hat{V}\sim({\bf 1},{\bf 1}, 0)$ & $V_{m}$ & $\lambda$ & $g^{\prime}$ \\
\hline
\end{tabular}
\end{center}
\caption{\small Particle content in the vector superfields in 
MSUSY331, where $a=1,2, \ldots ,8$.}
\label{gaugemssm}
\end{table}

We denote the $SU(3)_{L}$ gauge bosons by $V^{a}_{m}$ ($a=1,2, \ldots ,8$) and 
since 
\begin{eqnarray}
\left( {\bf 1}, {\bf 8}, 0 \right) \rightarrow 
\left( {\bf 1},{\bf 3}, 0 \right) \oplus 
\left( {\bf 1}, {\bf 2}, 3 \right) \oplus 
\left( {\bf 1}, {\bf 2}, -3 \right) \oplus 
\left( {\bf 1},{\bf 1}, 0 \right),
\end{eqnarray} 
we get a triplet, $V^{1}_{m},V^{2}_{m},V^{3}_{m}$, and also a 
singlet $V^{8}_{m}$ both with $Y=0$, plus the doublet of bileptons, 
$\left( V^{4}_{m},V^{5}_{m} \right)$, its hypercharge is 
$Y= 3$ and $\left( V^{6}_{m},V^{7}_{m} \right)$ and $Y=-3$. The charged gauge bosons of 
this model are \cite{mcr}
\begin{eqnarray}
W^{ \pm}_{m}(x)&=&-\frac{1}{\sqrt{2}}(V^{1}_{m}(x) 
\mp i V^{2}_{m}(x)), \,\
V^{ \pm}_{m}(x)=-\frac{1}{\sqrt{2}}(V^{4}_{m}(x) 
\pm i V^{5}_{m}(x)) \nonumber \\
U^{\pm \pm}_{m}(x) &=&- \frac{1}{\sqrt{2}}(V^{6}_{m}(x) 
\pm i V^{7}_{m}(x)).
\label{defcarbosons}
\end{eqnarray} 
The neutral gauge bosons are the photon
\begin{eqnarray}
A_{m}&=& \sin \theta_{W} \left( V^{3}_{m}- \sqrt{3}V^{8}_{m} \right) + 
\sqrt{1-4 \sin^{2} \theta_{W}}V_{m},
\end{eqnarray}
where $\theta_{W}$ is the Weinberg angel and we have also two massives neutral bosons defined as
\begin{eqnarray}
Z_{m}&=& \cos \theta_{W}V^{3}_{m}+ \sqrt{3}\tan \theta_{W}\sin \theta_{W}V^{8}_{m}- 
\tan \theta_{W}\sqrt{1-4 \sin^{2} \theta_{W}}V_{m}, \nonumber \\
Z^{\prime}_{m}&=& \frac{1}{\cos \theta_{W}}\left[
\sqrt{1-4 \sin^{2} \theta_{W}}V^{8}_{m}+ \sqrt{3}\sin \theta_{W}V_{m} \right].
\end{eqnarray}

In an recent article we explained the 
$W$-boson mass, presented by the CDF\footnote{Compare with 
Eq.(\ref{gaugemassSusygelmini}).}, in the MSUSY331 \cite{Rodriguez:2022hsj}
\begin{eqnarray}
M^{2}_{W^{\pm}}&=& \frac{g^{2}}{4} \left[
v^{2}_{SM} + v^{2}_{\sigma^{\prime}_{2}} +
2 \left( v^{2}_{\sigma_{1}}+v^{2}_{\sigma^{\prime}_{1}} \right)  \right], \nonumber \\
v^{2}_{SM}&=&v^{2}_{ \eta}+v^{2}_{ \rho}+v^{2}_{ \sigma_{2}}+
v^{2}_{ \eta^{\prime}}+v^{2}_{ \rho^{\prime}}, \nonumber \\
M^{2}_{V}&=& \frac{g^{2}}{4}\left[
v^{2}_{ \eta}+v^{2}_{ \chi}+v^{2}_{ \sigma_{2}}+v^{2}_{ \eta^{\prime}}+v^{2}_{ \chi^{\prime}}+
v^{2}_{\sigma^{\prime}_{2}}+ 
2 \left( v^{2}_{\sigma_{1}}+v^{2}_{\sigma^{\prime}_{1}} \right)  \right], \nonumber \\
M^{2}_{U}&=& \frac{g^{2}}{4} \left(
v^{2}_{\rho}+v^{2}_{\chi}+4v^{2}_{\sigma_{2}}+
v^{2}_{\rho^{\prime}}+v^{2}_{\chi^{\prime}}+
4v^{2}_{\sigma^{\prime}_{2}} 
\right), \nonumber \\
M^{2}_{Z^{0}}&\approx& \frac{1}{2} 
\left( \frac{g^{2}+4g^{\prime 2}}{g^{2}+3g^{\prime 2}} \right)
\left[ v^{2}_{SM}+ v^{2}_{\sigma^{\prime}_{2}}+ 
4 \left( v^{2}_{\sigma_{1}}+v^{2}_{\sigma^{\prime}_{1}} \right)  \right], \nonumber \\
M^{2}_{Z^{\prime}}&\approx& \frac{2}{3} (g^{2}+3g^{\prime 2}) 
(v^{2}_{ \chi}+v^{2}_{ \chi^{\prime}}), \nonumber \\
\rho &=& \frac{1+R}{1+R^{\prime}} \simeq 1+R-R^{\prime}
=1-2 \left( 
\frac{v^{2}_{\sigma_{1}}+v^{2}_{\sigma^{\prime}_{1}}}{v^{2}_{SM}} 
\right), \nonumber \\
R&=&\frac{v^{2}_{\sigma^{\prime}_{2}}}{v^{2}_{\rm SM}}+ 2 \left( \frac{v^{2}_{\sigma_{1}}+v^{2}_{\sigma^{\prime}_{1}}}{v^{2}_{SM}} 
\right),  \,\
R^{\prime}=\frac{v^{2}_{\sigma^{\prime}_{2}}}{v^{2}_{\rm SM}}+ 
4 \left( 
\frac{v^{2}_{\sigma_{1}}+v^{2}_{\sigma^{\prime}_{1}}}{v^{2}_{SM}} 
\right).
\label{rhoexpmsusy331}
\end{eqnarray} 
Those result is similar with our Eq.(\ref{gaugemassSusygelmini}). We 
present some phenomenological results in \ref{sec:fenomsusy331}.

The superpotencial in MSUSY331 is defined as \cite{331susy1,mcr,Rodriguez:2005jt}
\begin{equation}
W=W_{2}+ W_{3}+hc, 
\label{sp1}
\end{equation}
where each term above is defined as follows
\begin{eqnarray}
W_{2}&=&\sum_{i=1}^{3} \mu_{0i}(\hat{L}_{i}\hat{\eta}^{\prime})+ 
\mu_{ \eta} (\hat{\eta}\hat{\eta}^{\prime})+
\mu_{ \rho} (\hat{\rho}\hat{\rho}^{\prime})+ 
\mu_{ \chi} (\hat{\chi}\hat{\chi}^{\prime})+
\mu_{S} Tr[(\hat{S} \hat{S}^{\prime})]. 
\end{eqnarray} 
All the coefficients in the equation above have mass dimension. The last term of the 
superpotential in MSUSY331 has the following form
\begin{eqnarray}
W_{3}&=&\sum_{i=1}^{3} \left[ \sum_{j=1}^{3} \left( 
\sum_{k=1}^{3}\lambda_{1ijk} 
(\epsilon \hat{L}_{i}\hat{L}_{j}\hat{L}_{k})+
\lambda_{2ij} (\epsilon \hat{L}_{i}\hat{L}_{j}\hat{\eta})+
\lambda_{3ij} 
(\hat{L}_{i}\hat{S}\hat{L}_{j}) \right)+
\lambda_{4i}(\epsilon\hat{L}_{i}\hat{\chi}\hat{\rho})
\right] \nonumber \\
&+&
f_{1} (\epsilon \hat{\rho} \hat{\chi}\hat{\eta})+
f_{2} (\hat{\chi}\hat{S}\hat{\rho})  +
f_{3} (\hat{\eta}\hat{S}\hat{\eta}) +
f_{4}\epsilon_{ijk}\epsilon_{lmn}\hat{S}_{il}\hat{S}_{jm}\hat{S}_{kn}+
f^{\prime}_{1} (\epsilon \hat{\rho}^{\prime}\hat{\chi}^{\prime} \hat{\eta}^{\prime})+
f^{\prime}_{2} (\hat{\chi}^{\prime} 
\hat{S}^{\prime}\hat{\rho}^{\prime})  
\nonumber \\ &+&
f^{\prime}_{3} (\hat{\eta}^{\prime} 
\hat{S}^{\prime}\hat{\eta}^{\prime}) +
f^{\prime}_{4}\epsilon_{ijk}\epsilon_{lmn}
\hat{S}^{\prime}_{il}\hat{S}^{\prime}_{jm}
\hat{S}^{\prime}_{kn} +
\sum_{i=1}^{3} \left[ \kappa_{1i}
(\hat{Q}_{3} \hat{\eta}^{\prime}) \hat{u}^{c}_{i}+ 
\kappa_{2i} 
(\hat{Q}_{3} \hat{\rho}^{\prime}) \hat{d}^{c}_{i} \right]+
\kappa_{3} 
(\hat{Q}_{3}\hat{\chi}^{\prime}) 
\hat{J}^{c} 
\nonumber \\
&+&
\sum_{\alpha =1}^{2} \left[
\sum_{i=1}^{3} \left(
\kappa_{4\alpha i} 
(\hat{Q}_{\alpha} \hat{\eta}) \hat{d}^{c}_{i}+
\kappa_{5\alpha i} 
(\hat{Q}_{\alpha} \hat{\rho}) \hat{u}^{c}_{i} \right) +
\sum_{\beta =1}^{2} 
\kappa_{6\alpha\beta}
(\hat{Q}_{\alpha} \hat{\chi}) \hat{j}^{c}_{\beta} 
+ \sum_{i=1}^{3}\sum_{j=1}^{3}
\kappa_{7\alpha ij} 
(\hat{Q}_{\alpha} \hat{L}_{i}) 
\hat{d}^{c}_{j} \right]
\nonumber \\
&+&
\sum_{i=1}^{3} \left\{ \sum_{j=1}^{3}\left[ \sum_{k=1}^{3}
\xi_{1ijk} \hat{d}^{c}_{i} \hat{d}^{c}_{j} \hat{u}^{c}_{k}+
\sum_{\beta =1}^{2} \left(
\xi_{2ij\beta} \hat{u}^{c}_{i} \hat{u}^{c}_{j} \hat{j}^{c}_{\beta}
\right]+
\xi_{3 i\beta} \hat{d}^{c}_{i} \hat{J}^{c} \hat{j}^{c}_{\beta} \right) 
\right\}. 
\label{sp3m1}
\end{eqnarray}
All the coefficients in $W_{3}$ are dimensionless. Note that the scalar 
$S^{\prime}$ does not have any type of coupling with the fermions of the model, 
so the VEV of its neutral components can be a few GeV, it means our scalar 
field $S^{\prime}$ is similar to the scalar field $\Delta_{2}$ in the 
SUSYGR, see our Eq.(\ref{suppotSUSYGR}).

We want to emphasize, the gauge bosons $V^{\pm}$ and $U^{\pm \pm}$ are 
bileptons, see our Eq.(\ref{lq}). The term $(\hat{L}\hat{S}\hat{L})$ will 
introduce a term similar to Eq.(\ref{yukawatermogelmini}) then some of our 
scalars are also bilepton. The terms proportional for 
$f_{2},f_{4},f^{\prime}_{2}$ and $f^{\prime}_{4}$\footnote{They are similar to 
the coefficients $\lambda_{1}$ and $\lambda_{2}$ in the SUSYGR, see our 
Eq.(\ref{suppotSUSYGR}).} violate the conservation of the quantum number
\begin{equation}
{\cal F} \equiv B+L,
\label{fquantumnumber}
\end{equation}
where $B$ is the baryon number while $L$ is the total lepton number
\footnote{See our Eqs.(\ref{accidentalsyminSM},\ref{potmajoronmp})}
\cite{331susy1,mcr,Rodriguez:2005jt,Pleitez:1992xh}. However, if we assume the 
global $U(1)_{\cal F}$ symmetry, it allows us to introduce the $R$-conserving 
symmetry, in a similar way as we define $R$-parity in MSSM 
\cite{Rodriguez:2019mwf}, defined as \cite{Rodriguez:2010tn}
\begin{equation}
R \equiv (-1)^{2S+3{\cal F}}.
\end{equation}
Where $S$ is the spin for a given state. The ${\cal F}$ number attribution is
\begin{equation}
\begin{array}{c}
{\cal F}(U^{--})={\cal F}(V^{-}) = - {\cal F}(J)= {\cal F}(j_{1,2})=
{\cal F}(\rho^{--})  \\  
= {\cal F}(\chi^{--}) ={\cal F}(\chi^{-}) = 
{\cal F}(\eta^{-}_{2})={\cal F}(\sigma_{1}^{0})=2.
\end{array}
\label{efe}
\end{equation} 
For instance, if we allow the $\xi_{1}$ term it implies in proton 
decay~\cite{pal2}. We will show below those terms will allow us to implement 
the scheme of Ma, see our Eq.(\ref{potmajoronmp}), in the MSUSY331 in the same 
way as we implement this mechanism in SUSYGR as we discuss after our 
Eq.(\ref{suppotSUSYGR}).

The soft terms to break SUSY\footnote{We are not considering the Gaugino 
Mass terms neither sleptons and squarks, these soft terms are defined in  
\cite{Rodriguez:2019mwf}.} is written as
\begin{eqnarray}
{\cal L}^{soft}_{scalar}&=&-
m^{2}_{ \eta}(\eta^{ \dagger}\eta)-
m^{2}_{ \rho}(\rho^{ \dagger}\rho)-
m^{2}_{ \chi}(\chi^{ \dagger}\chi)-
m^{2}_{S}Tr[(S^{ \dagger}S)]
-m^{2}_{\eta^{\prime}}
(\eta^{\prime \dagger}\eta^{\prime})-
m^{2}_{\rho^{\prime}}
(\rho^{\prime \dagger}\rho^{\prime})
\nonumber \\ &-&
m^{2}_{\chi^{\prime}}
(\chi^{\prime \dagger}\chi^{\prime}) 
-m^{2}_{S^{\prime}}
Tr[(S^{\prime \dagger} S^{\prime})]
+[k_{1}(\epsilon \rho \chi \eta)+
k_{2}(\eta S \eta)+
k_{3}(\chi S \rho)+
k_{4} \epsilon \epsilon SSS
\nonumber \\ 
&+&
k^{\prime}_{1}(\epsilon \rho^{\prime}\chi^{\prime} \eta^{\prime})+
k^{\prime}_{2}(\eta^{\prime}S^{\prime}\eta^{\prime})+
k^{\prime}_{3}(\chi^{\prime}S^{\prime} \rho^{\prime})+
k^{\prime}_{4} \epsilon \epsilon S^{\prime}S^{\prime}S^{\prime}+
hc], 
\end{eqnarray}
$k_{1,2,3,4},k^{\prime}_{1,2,3}$ and $k^{\prime}_{4}$ has mass dimension. 

The pattern of the symmetry breaking in MSUSY331 is given by the following scheme
\begin{eqnarray}
&\mbox{MSUSY331}&
\stackrel{{\cal L}_{soft}}{\longmapsto}
\mbox{SU(3)}_C\ \otimes \ \mbox{SU(3)}_{L}\otimes \mbox{U(1)}_{N}
\stackrel{\langle\chi\rangle \langle \chi^{\prime}\rangle}{\longmapsto}
\mbox{SU(3)}_{C} \ \otimes \ \mbox{SU(2)}_{L}\otimes
\mbox{U(1)}_{Y} \nonumber \\
&\stackrel{\langle \rho, \eta, S \rho^{\prime},\eta^{\prime},S^{\prime}\rangle}{\longmapsto}&
\mbox{SU(3)}_{C} \ \otimes \ \mbox{U(1)}_{em}.
\label{breaksusy331tou1}
\end{eqnarray}
When $\sigma_{2}$ get VEV the charged leptons obtain their masses. The fields 
$\sigma^{\prime}_{2}, \sigma_{1}$ and $\sigma^{\prime}_{1}$ are useful to 
explain the $W$-boson mass presented by the CDF in the MSUSY331 in two possible 
ways \cite{Rodriguez:2022hsj}
\begin{itemize}
\item[1-)] $v_{\sigma_{1}}=v_{\sigma^{\prime}_{1}}=0$ GeV and 
$v_{\sigma_{2}}=10$ GeV and
$v_{\sigma^{\prime}_{2}}=11.19$ GeV.
\item[2-)] We can also choose 
\begin{equation}
v_{\sigma_{1}}\neq 0, v_{\sigma^{\prime}_{1}}\neq 0, 
v_{\sigma^{\prime}_{2}}\neq 0.
\label{defcase2}
\end{equation}
In this case those VEV have to satisfy the following
\begin{equation}
v_{\sigma^{\prime}_{1}}= \sqrt{\frac{2\delta M^{2}_{W}}{g}- 
\frac{v^{2}_{\sigma^{\prime}_{2}}}{2}-v^{2}_{\sigma_{1}}}, \,\ 
v_{\sigma^{\prime}_{2}}=
8.7691841 \,\ GeV,
\label{sigma1VSsigma1p}
\end{equation}
some possible values is shown in our Tab.(\ref{vs1Xvs1p}). We will 
use the following point for our 
numerical analyses $v_{\sigma_{1}}=4.89$ GeV, 
$v_{\sigma_{2}}=10$ GeV,
$v_{\sigma^{\prime}_{1}}=0.5$ GeV and
$v_{\sigma^{\prime}_{2}}=8.77$ GeV.
\end{itemize}
We show some details of our scalar potential in 
Apps.(\ref{psusy1},\ref{constpot}).

\begin{table}
\begin{tabular}{|c|c|}
\hline
$v_{\sigma_{1}}$ (GeV) & $v_{\sigma^{\prime}_{1}}$ (GeV) \\ 
\hline
0.5 & 4.89453  \\
1.0 & 4.81730  \\
1.5 & 4.68577  \\
2.0 & 4.49515  \\
2.5 & 4.2375  \\
3.0 & 3.89954  \\
3.5 & 3.4578  \\
4.0 & 2.86468  \\
4.5 & 1.98907  \\
\hline
\end{tabular}
\caption{Defined some values of $v_{\sigma_{1}}$ we show the numerical values for $v_{\sigma^{\prime}_{1}}$ GeV take of our study presented in \cite{Rodriguez:2022hsj}.}
\label{vs1Xvs1p}
\end{table}

\section{Case 1}

This case is defined by the following choice
\begin{equation}
v_{\sigma_{1}}=v_{\sigma^{\prime}_{1}}=0 \,\ GeV. 
\end{equation}
We will use below the following set of parameters in the scalar potential
\begin{eqnarray}
f_{1}=1.12, \quad f_{3}=1.03,\quad f^{\prime}_{1}=f^{\prime}_{3}=10^{-6},
\quad {\rm (dimensionless)}
\label{fs}
\end{eqnarray}
and 
\begin{eqnarray}
-k_{1}=k^{\prime}_{1}=10,\;k_{3}=k^{\prime}_{3}=-100,\;  
-\mu_{\eta}=\mu_{\rho}=- \mu_{s}= \mu_{\chi}=1000 \quad GeV, 
\nonumber \\
\label{ks}
\end{eqnarray}
we also use the constraint\footnote{To recover the VEV of SM given in our 
Eqs.(\ref{vevsm},\ref{rhoexpmsusy331}).} 
\begin{equation}
v^{2}_{\eta}+ v^{2}_{\eta^{\prime}}+v^{2}_{\rho}+v^{2}_{\rho^{\prime}}+
2(v^{2}_{\sigma^{0}_{2}}+v^{2}_{\sigma^{\prime}_{2}})=(246\; GeV)^{2}.
\end{equation} 
Assuming that 
\begin{eqnarray}
v_{\eta}&=&20 \,\ GeV, \,\  
v_{\chi}=1000 \,\ GeV, \,\ 
v_{\sigma_{2}}=10 \,\ GeV, \nonumber \\ 
v_{\eta^{\prime}}&=&v_{\rho^{\prime}}=v_{\chi^{\prime}}=1 \,\ GeV,
\label{vevmassspectrum}
\end{eqnarray}  
the value of $v_{\rho}$ is fixed by the constraint above. We get similar 
values are presented in \cite{Rodriguez:2005jt}. 

We define the following basis for the neutral scalar sector (CP even)
\begin{equation}
\left(
H_{\eta},H_{\rho},H_{\chi},H_{\sigma^{0}_{2}},H_{\eta^{\prime}},H_{\rho^{\prime}},
H_{\chi^{\prime}},H_{\sigma^{\prime 0}_{2}}
\right) .
\end{equation} 
All these fields are massive and we represent each mass value as $M_{H_{j}}>M_{H_{i}}$, 
$i=1,\cdots,8$, and  with $j>i$. Our mass spectrum in (GeV) for the VEV given in our 
Eq.(\ref{vevmassspectrum}) is as follows
\begin{eqnarray}
M_{H^{0}_{1}}&\equiv& M_{h^{0}}=125.577, \,\ 
M_{H^{0}_{2}}\equiv M_{H^{0}}=411.850, \,\ 
M_{H^{0}_{3}}=633.157, \nonumber \\ 
M_{H^{0}_{4}}&=&1017.34, \,\ 
M_{H^{0}_{5}}=1193.52, \,\ 
M_{H^{0}_{6}}=1254.76, \,\ 
M_{H^{0}_{7}}=1375.39, \nonumber \\ 
M_{H^{0}_{8}}&=&4791.17.
\end{eqnarray}

\begin{figure}[ht]
\begin{center}
\vglue -0.009cm
\mbox{\epsfig{file=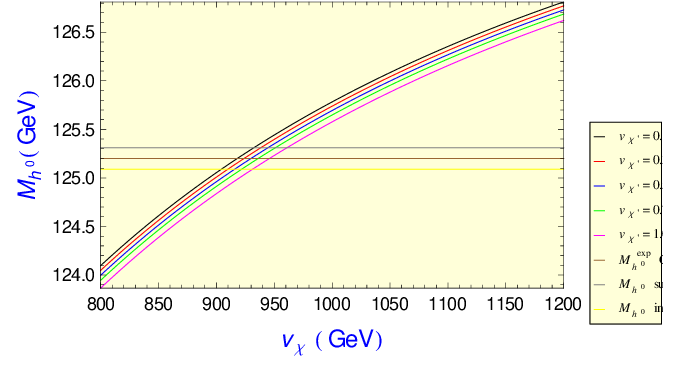,width=0.7\textwidth,angle=0}}       
\end{center}
\caption{We show the prediction about the behaviour of the masses of lightest 
scalars, $M_{h^{0}}$, in terms of $v_{\chi}$ for several values of 
$v_{\chi^{\prime}}$ show in the box in several colors in case 1. 
$M^{exp}_{h^{0}}$ (straight line Brown) is the 
central experimental values given at our Eq.(\ref{expvalh}), 
$M^{sup}_{h^{0}}=M^{exp}_{h^{0}}+0.11$ (straight line gray) while 
$M^{inf}_{h^{0}}=M^{exp}_{h^{0}}-0.11$ (straight line yellow).}
\label{fig1h}
\end{figure}

\begin{figure}[ht]
\begin{center}
\vglue -0.009cm
\mbox{\epsfig{file=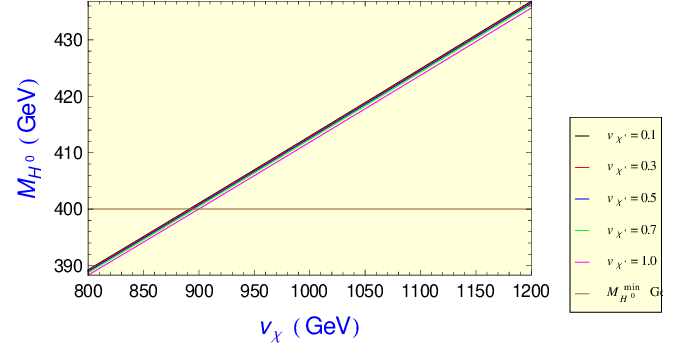,width=0.7\textwidth,angle=0}}       
\end{center}
\caption{We show the prediction about the behaviour of the masses of the usual 
heavy scalar, $M_{H^{0}}$, in terms of $v_{\chi}$ for several values of 
$v_{\chi^{\prime}}$ show in the box in several colors in case 1. 
$M^{exp}_{H^{0}}$ is the lower experimental values, given in our 
Eq.(\ref{limATLASH}), (straight line Brown).}
\label{fig1H}
\end{figure} 

In the neutral imaginary scalar (pseudoscalar) sector (CP odd) we define the 
following base
\begin{equation}
\left(
F_{\eta},F_{\rho},F_{\chi},F_{\sigma^{0}_{2}},F_{\eta^{\prime}},F_{\rho^{\prime}},
F_{\chi^{\prime}},F_{\sigma^{\prime 0}_{2}}
\right) ,
\end{equation} 
and we have verified analytically, we have two Goldstones bosons 
($G^{0}_{Z^{0}},G^{0}_{Z^{\prime 0}}$). These Goldstones bosons will give 
masses for the neutral gauge boson $Z^{0}$ and $Z^{\prime 0}$, respectively. 
The other six physical pseudoscalars  have the following masses in (GeV) when we use 
the VEV defined in our Eq.(\ref{vevmassspectrum}), they are the following values 
\begin{eqnarray}
M_{A^{0}_{1}}&\equiv& M_{A^{0}}=411.378, \,\ 
M_{A^{0}_{2}}=633.183, \,\ 
M_{A^{0}_{3}}=1017.31, \,\ 
M_{A^{0}_{4}}=1193.52, \nonumber \\ 
M_{A^{0}_{5}}&=&1375.39, \,\ 
M_{A^{0}_{6}}=4791.17.
\end{eqnarray}

\begin{figure}[ht]
\begin{center}
\vglue -0.009cm
\mbox{\epsfig{file=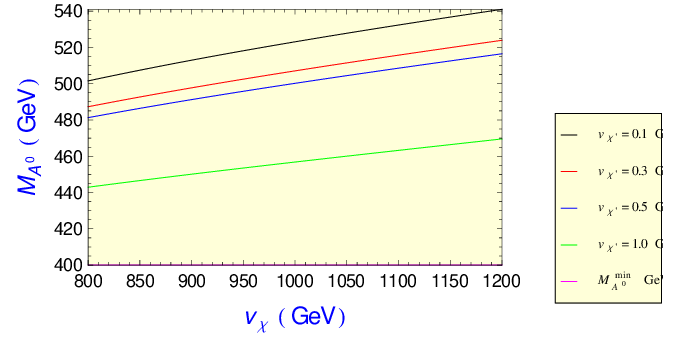,width=0.7\textwidth,angle=0}}       
\end{center}
\caption{We show the prediction about the behaviour of the masses of the lightest 
pseudoscalar, $M_{A^{0}}$, in terms of $v_{\chi}$ for several values of 
$v_{\chi^{\prime}}$ show in the box in several colors 
in case 1. $M^{exp}_{A^{0}}$ is the lower experimental values, given in our 
Eq.(\ref{limATLASA}), (straight line Brown).}
\label{fig1A}
\end{figure}

\begin{figure}[ht]
\begin{center}
\vglue -0.009cm
\mbox{\epsfig{file=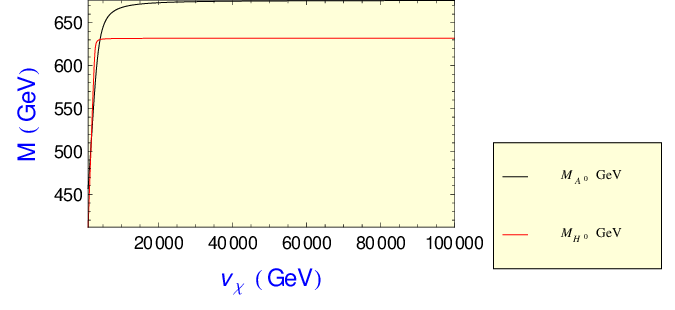,width=0.7\textwidth,angle=0}}       
\end{center}
\caption{We show the prediction about the behaviour of the masses of the lightest 
pseudoscalar, $M_{A^{0}}$ and $M_{H^{0}}$, in terms of $v_{\chi}$ for 
$v_{\chi^{\prime}}=1$ GeV.}
\label{fig1AH}
\end{figure}

Possible values of the masses of scalars $h^{0}, H^{0}$ and pseudoscalar $A^{0}$ 
in terms of $v_{\chi}$ are show in our Figs.(\ref{fig1h},\ref{fig1H},\ref{fig1A}), 
respectively. To obtain upper limits on the masses of $H^{0}$ and $A^{0}$ we draw 
Fig.(\ref{fig1AH}), where we show $M_{H^{0}}$ and $M_{A^{0}}$ for $v_{\chi}>20$ TeV. 
From these figures we can say that
\begin{eqnarray}
900 \,\ GeV &\leq& v_{\chi} \leq 1000 \,\ GeV, \,\ 
M_{h^{0}}\approx 125.20 \,\ GeV; \,\ 
M_{H^{0}}>M_{A^{0}} ; \nonumber \\ 
M_{H^{0}}&<& 632 \,\ GeV; \,\ 
M_{A^{0}}< 676 \,\ GeV.
\end{eqnarray}
Therefore, our results for the masses of neutral scalars are within current 
experimental limits show in our 
Eqs.(\ref{expvalh},\ref{limATLASH},\ref{limATLASA}).

In our opinion, it is interesting to do a phenomenological study regarding the decays 
of the scalars $h^{0}$, $H^{0}$ and pseudoscalar $A^{0}$ in MSUSY331. A first 
series studying the decay of scalars in the 331 model without SUSY was 
presented in \cite{Okada:2016whh,Fan:2022dye}.

It would be, also, so interesting to diagonalize these matrices in 
details. It means, we must to study the perturbative and unitary conditions for 
the potential; verify the Vacuum stability and of course the minimization 
condition. In our opinion, those analyses must be done first in the (m331) and 
then make in the MSUSY331. We have previously done a similar analysis in\footnote{I am still in debt with Prof. Aahron Davidson, who introduced me to this 
interesting model, which should be better studied.}  \cite{Rodriguez:2021rul}.

\begin{figure}[ht]
\begin{center}
\vglue -0.009cm
\mbox{\epsfig{file=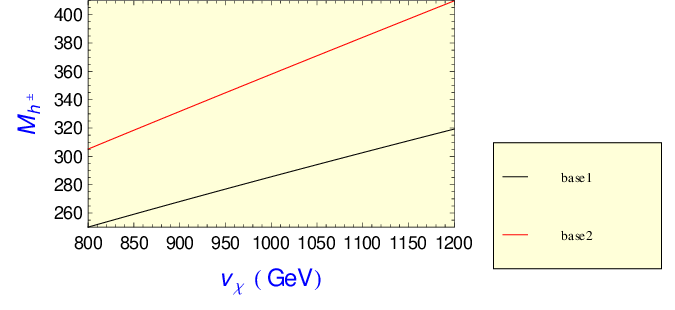,width=0.7\textwidth,angle=0}}       
\end{center}
\caption{We show the prediction about the behaviour of the masses 
$\left( M_{h^{\pm}} \right)_{b1}$ (straight line black) 
and $\left( M_{h^{\pm}} \right)_{b2}$ (straight line red) in terms of 
$v_{\chi}$ when $v_{\chi^{\prime}}=1$ GeV in case 1.}
\label{fig1hpb1-b2}
\end{figure}

In the Single charged Scalars we have two Goldstones bosons $G^{\pm}_{W^{\pm}}$ and 
$G^{\pm}_{V^{\pm}}$. They will give masses for the gauge bosons $W^{\pm}$ and $V^{\pm}$, 
respectively. And we also have nine massive states in two unrelated bases. 
The first base is defined as
\begin{eqnarray}
\left(
\eta^{-}_{1},\rho^{-},\eta^{\prime -}_{1},\rho^{\prime -},h^{-}_{1},
h^{\prime -}_{1}
\right)^{T},
\label{base1single}
\end{eqnarray} 
we get the following masses in (GeV) for the VEV given in our 
Eq.(\ref{vevmassspectrum})
\begin{eqnarray}
(M_{H^{\pm}_{1}})_{b1}&\equiv& \left( M_{h^{\pm}} \right)_{b1}= 287.399, \,\ 
(M_{H^{\pm}_{2}})_{b1}= 970.554, \,\ 
(M_{H^{\pm}_{3}})_{b1}= 1155.81, \nonumber \\ 
(M_{H^{\pm}_{4}})_{b1}&=& 1343.41, \,\ 
(M_{H^{\pm}_{5}})_{b1}= 4781.44.
\end{eqnarray} 
The second base is defined as
\begin{eqnarray}
\left(
\eta^{-}_{2},\chi^{-},\eta^{\prime -}_{2},\chi^{\prime -},h^{-}_{2},
h^{\prime -}_{2}
\right)^{T},
\label{base2single}
\end{eqnarray}
for this new basis the eigenvalues ​​(GeV) have the following values for the VEV 
given in our Eq.(\ref{vevmassspectrum})
\begin{eqnarray}
(M_{H^{\pm}_{1}})_{b2}&\equiv& \left( M_{h^{\pm}} \right)_{b2}= 357.111, \,\ 
(M_{H^{\pm}_{2}})_{b2}= 645.823, \,\ 
(M_{H^{\pm}_{3}})_{b2}= 931.247, \nonumber \\ 
(M_{H^{\pm}_{4}})_{b2}&=& 1175.67, \,\ 
(M_{H^{\pm}_{5}})_{b2}= 4781.45.
\end{eqnarray} 
From these data we realize that $\left( M_{h^{\pm}} \right)_{b2}$ is heavier than 
$\left( M_{h^{\pm}} \right)_{b1}$, which we show in more detail in our 
Fig.(\ref{fig1hpb1-b2}) for several values of $v_{\chi}$ when 
$v_{\chi^{\prime}} =1$ GeV. The masses values in this sector are 
in agreement with the actual experimental data.

\begin{figure}[ht]
\begin{center}
\vglue -0.009cm
\mbox{\epsfig{file=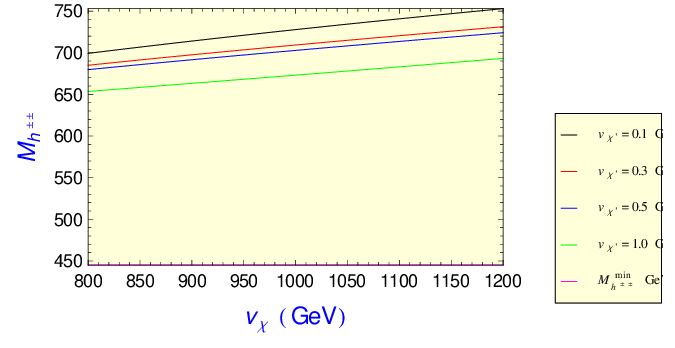,width=0.7\textwidth,angle=0}}       
\end{center}
\caption{We show the prediction about the behaviour of the masses of $M_{h^{\pm \pm}}$ 
in terms of $v_{\chi}$ for some values of $v_{\chi^{\prime}}$ show 
in the box in several colors.}
\label{fig1hpp}
\end{figure}

In the Double charged Scalars we have one Goldstone boson, 
$G^{\pm \pm}_{U^{\pm \pm}}$, to give mass for the gauge boson $U^{\pm \pm}$ and seven 
massive states, and their masses in (GeV) for the VEV given in our 
Eq.(\ref{vevmassspectrum}) are 
\begin{eqnarray}
M_{H^{\pm \pm}_{1}}&\equiv& M_{h^{\pm \pm}}=530.175, 
M_{H^{\pm \pm}_{2}}= 813.436, \,\ 
M_{H^{\pm \pm}_{3}}= 1058.42, \nonumber \\ 
M_{H^{\pm \pm}_{4}}&=& 1123.61, \,\ 
M_{H^{\pm \pm}_{5}}= 1240.58, \,\ 
M_{H^{\pm \pm}_{6}}=1277.02, \nonumber \\ 
M_{H^{\pm \pm}_{7}}&=& 1396.67.
\end{eqnarray} 
The mass of the lightest state as function of $v_{\chi}$ is shown in our 
Fig.(\ref{fig1hpp}) and they are in agreement with the actual experimental data. The mass hierarchi in this case is
\begin{equation}
M_{h^{\pm \pm}}> \left( M_{h^{\pm}} \right)_{b2}>
M_{A^{0}}>M_{H^{0}}>\left( M_{h^{\pm}} \right)_{b1}.
\end{equation}

\section{Case 2}

This case is defined by Eq.(\ref{defcase2}). Now in the sector of neutral scalar, we 
have two $10 \times 10$. The new basis for the even CP case is
\begin{eqnarray}
\left(  H_{\eta},H_{\rho},H_{\chi},H_{\sigma_{1}},H_{\sigma_{2}},
H_{\eta^{\prime}},H_{\rho^{\prime}},H_{\chi^{\prime}},
H_{\sigma^{\prime}_{1}},H_{\sigma^{\prime}_{2}} \right)^{T},
\end{eqnarray}
while the base for CP odd is
\begin{eqnarray}
\left(  F_{\eta},F_{\rho},F_{\chi},F_{\sigma_{1}},F_{\sigma_{2}},
F_{\eta^{\prime}},F_{\rho^{\prime}},F_{\chi^{\prime}},
F_{\sigma^{\prime}_{2}} \right)^{T}.
\end{eqnarray} 
For the single Charged scalars instead of having two distinct bases we have a single 
one, as we will present below.

If we suppose  $f_{2}=f_{4}=f^{\prime}_{2}=f^{\prime}_{4}=0$, we obtain three 
Goldstone bosons in the CP odd sector, that is, one more Goldstone than 
necessary\footnote{Remember $v_{\sigma^{0}_{1}}$ break $L$ symmetry as 
$V_{\Delta}$ do the same in Gelmini-Roncadelli scheme and 
$V_{\Delta_{1}},V_{\Delta_{2}}$ in SUSYGR.}. 
This means we have, one Goldstone boson $G^{0}_{Z^{0}}$ for give mass for 
$Z^{0}$, another one $G^{0}_{Z^{\prime 0}}$ for give mass for $Z^{\prime 0}$ 
and the last one is the Majoron $M^{0}$. 

We can remove this Majoron, in similar way as done in our 
Eq.(\ref{potmajoronmp}), when we impose that the parameters 
$f_{2}$, $f_{4}$, $f^{\prime}_{2}$ and $f^{\prime}_{4}$ are different from 
zero. In other words, we see that these terms 
allow us to satisfactorily implement scheme of Ma, see our 
Eq.(\ref{potmajoronmp}), in MSUSY331 in a similar way as implemented in SUSYGR, 
see our Eqs(\ref{suppotSUSYGR},\ref{sp3m1}).

We will brief present our results when $v_{\chi}=1000$ GeV and 
$v_{\chi^{\prime}}=1$ GeV and we also use the following numerical, in GeV, parameters
\begin{eqnarray}
f_{2}&=&f_{4}=1.11, \,\  
f^{\prime}_{2}=f^{\prime}_{4}=10^{-6}, \nonumber \\
-k_{2}&=&k^{\prime}_{2}=10,\;k_{4}=k^{\prime}_{4}=-100.
\label{ks}
\end{eqnarray}
\begin{itemize}
\item CP even we get and the mass spectrum in (GeV) is as follows
\begin{eqnarray}
M_{H^{0}_{1}}&\equiv& M_{h^{0}}= 125.262, \,\ 
M_{H^{0}_{2}}\equiv M_{H^{0}}= 422.021, \,\ 
M_{H^{0}_{3}}= 491.822, \nonumber \\ 
M_{H^{0}_{4}}&=& 674.681, \,\ 
M_{H^{0}_{5}}= 1052.59, \,\ 
M_{H^{0}_{6}}= 1200.45, \,\ 
M_{H^{0}_{7}}= 1281.98, \nonumber \\ 
M_{H^{0}_{8}}&=& 1369.16, \,\ 
M_{H^{0}_{9}}= 1411.46, \,\ 
M_{H^{0}_{10}}= 4777.62;
\end{eqnarray} 
\item CP odd the base have the following masses, with the same parameters 
as before, in (GeV) 
\begin{eqnarray}
M_{A^{0}_{1}}&\equiv& M_{A^{0}}= 420.963, \,\ 
M_{A^{0}_{2}}= 495.397, \,\ 
M_{A^{0}_{3}}= 674.74, \nonumber \\ 
M_{A^{0}_{4}}&=& 1052.56, \,\ 
M_{A^{0}_{5}}= 1200.45, \,\ 
M_{A^{0}_{6}}= 1369.17, \,\ 
M_{A^{0}_{7}}= 1411.42, \nonumber \\ 
M_{A^{0}_{8}}&=& 4777.61;
\end{eqnarray}
\item Single charged Scalars we get the following masses, in (GeV)
\begin{eqnarray}
M_{H^{\pm}_{1}}&\equiv& M_{h^{\pm}}= 299.376, \,\ 
M_{H^{\pm}_{2}}= 334.292, \,\ 
M_{H^{\pm}_{3}}= 440.198, \nonumber \\ 
M_{H^{\pm}_{4}}&=& 632.72, \,\ 
M_{H^{\pm}_{5}}= 931.934, \,\ 
M_{H^{\pm}_{6}}= 967.829, \,\ 
M_{H^{\pm}_{7}}= 1307.3, \nonumber \\ 
M_{H^{\pm}_{8}}&=& 1371.22, \,\ 
M_{H^{\pm}_{9}}= 4731.05, \,\ 
M_{H^{\pm}_{10}}= 4760.51;
\end{eqnarray}
\item Double charged Scalars we get the following masses, in (GeV)
\begin{eqnarray}
M_{H^{\pm \pm}_{1}}&\equiv& M_{h^{\pm \pm}}= 526.523, \,\ 
M_{H^{\pm \pm}_{2}}= 537.62, \,\ 
M_{H^{\pm \pm}_{3}}= 804.524, \nonumber \\ 
M_{H^{\pm \pm}_{4}}&=& 989.578, \,\ 
M_{H^{\pm \pm}_{5}}= 999.689, \,\ 
M_{H^{\pm \pm}_{6}}= 1247.99, \nonumber \\ 
M_{H^{\pm \pm}_{7}}&=& 1311.01.
\end{eqnarray} 
\end{itemize}

We also get the results show in our 
Tabs.(\ref{tabmhHAhhip3t1},\ref{tabmhHAhhip3t2}), where we notice 
the masses for $M_{h^{0}}$, $M_{H^{0}}$, $M_{A^{0}}$, $M_{h^{\pm}}$ and 
$M_{h^{\pm \pm}}$ are in agreement with the actual experimental values 
and we can conclude
\begin{itemize}
\item $M_{h^{0}}\sim 125.5$ GeV;
\item $M_{H^{0}}>400$ GeV;  $M_{A^{0}}>400$ GeV;
\item $M_{h^{\pm}}>M_{h^{\pm \pm}}>M_{H^{0}}>M_{A^{0}}$. 
\end{itemize}

\begin{table}
\begin{tabular}{|c|c|c|c|c|}
\hline 
$v_{\chi^{\prime}}$ (GeV) & $M_{h^{0}}$ (GeV)  & $M_{H^{0}}$ (GeV) & 
$M_{A^{0}}$ (GeV) & $M_{h^{\pm}}$ (GeV)  \\
\hline 
15.30 & 125.596 & 422.381  & 422.273  & 813.965 \\
15.35 & 125.483 & 422.373  & 422.271  & 813.930 \\
15.40 & 125.396 & 422.366  & 422.269  & 813.896 \\
15.45 & 125.309 & 422.358  & 422.267  & 813.861 \\
15.50 & 125.220 & 422.350  & 422.264  & 813.826 \\
\hline
\end{tabular}
\caption{The masses of $M_{h^{0}}, M_{H^{0}}, M_{A^{0}}$ and $M_{h^{\pm}}$ for 
$v_{\chi}=10$ TeV and some values of $v_{\chi^{\prime}}$ and under 
this suppositions we get $M_{h^{\pm \pm}}=527.140$ GeV.}
\label{tabmhHAhhip3t1}
\end{table}

\begin{table}
\begin{tabular}{|c|c|c|c|}
\hline 
$v_{\chi}$ (GeV) & $M_{h^{0}}$ (GeV)  & $M_{H^{0}}$ (GeV) & 
$M_{A^{0}}$ (GeV)   \\
\hline 
8500 & 124.871 & 422.349  & 422.223   \\
9000 & 124.996 & 422.349  & 422.245   \\
9500 & 125.112 & 422.350  & 422.255   \\
10000 & 125.220 & 422.350  & 422.264   \\
10500 & 125.323 & 422.351  & 422.271   \\
\hline
\end{tabular}
\caption{The masses of $M_{h^{0}}, M_{H^{0}}, M_{A^{0}}$ and  for 
$v_{\chi^{\prime}}=15.5$ GeV and some values of $v_{\chi^{\prime}}$ and under 
this suppositions we get $M_{h^{\pm}}813.826$ GeV and $M_{h^{\pm \pm}}=527.140$ GeV.}
\label{tabmhHAhhip3t2}
\end{table}

\section{Conclusions}
\label{sec:conclusion}
We have studied the scalar potential of supersymmetric 
$331$ model with all the VEV from the sextet and anti-sextet are not null 
and we show our lightest scalar fields has mass near the value 
$M_{h^{0}}\approx 125.5$ GeV, with concordance with the experimental values 
for Higgs masses given by Eq.(\ref{expvalh}).  We also show that in MSUSY331, we 
have $M_{A^{0}},M_{H^{0}}>400$ GeV, $M_{h^{\pm}}>181$ GeV and 
$M_{h^{\pm \pm}}>445$ GeV. Regarding which is heavier, $H^{0}$ or $A^{0}$, the 
answer depends on the condition. We show that in {\it case 1} we have 
$M_{A^{0}}>M_{H^{0}}$, see our Fig.(\ref{fig1AH}), and in {\it case 2} the 
opposite, that is, $M_{H^{0}}>M_{A^{0}}$ as we show in our 
Tab.(\ref{tabmhHAhhip3t2}). 

\begin{center}
{\bf Acknowledgments} 
\end{center}
We would like thanks V. Pleitez for useful discussions above $331$ models. 
We also to thanks IFT for the nice hospitality during my several visit 
to perform my studies about the $331$ models and also for done 
this article.

\appendix

\section{Some phenomenological Results in MSUSY331.}
\label{sec:fenomsusy331}

The interaction between the charged bosons with the leptons are given by \cite{mcr,Rodriguez:2010tn}
\begin{eqnarray}
{\cal L}_l^{CC}&=&-\frac{g}{\sqrt{2}}\sum_{l}\left(\bar{\nu}_{lL}\gamma^{m}V_{\rm PMNS}l_{L}W^{+}_{m}+ 
\bar{l}^{c}_{L}\gamma^{m}U_{V}\nu_{lL} V^{+}_{m}+
\bar{l}^{c}_{L}\gamma^{m} U_{U}l_{L}U^{++}_{m}+hc \right). \nonumber \\
\label{lq}
\end{eqnarray}
The $V_{\rm PMNS}$ is the Pontecorvo-Maki-Nakagawa-Sakata mixing matrix. 
There are new mixing matrices given by $U_{V}$ and $U_{U}$. 
The bosons $U^{--}$ and $V^{-}$ are called bileptons because they couple to 
two leptons; thus they have two units of lepton number, it means 
$L=L_{e}+L_{\mu}+L_{\tau}=2$. This model 
does not conserve separate family lepton number, $L_{e}$, $L_{\mu}$ and 
$L_{\tau}$ but only the total lepton number $L$ is conserved. We have already showed 
that in the M\o ller scattering and in muon-muon scattering we can show that 
left-right asymmetries $A_{RL}(ll)$  are very sensitive to a doubly charged vector 
bilepton resonance but they are insensitive to scalar ones 
\cite{Montero:1998ve,Montero:1999en,Montero:2000ch}.

Similarly, we have the neutral currents coupled to both $Z^{0}$ and
$Z^{\prime 0}$ massive vector bosons, according to the Lagrangian
\begin{equation}
{\cal L}_{\nu}^{NC}=-\frac{g}{2}\frac{M_{Z^{0}}}{M_{W^{\pm}}}
\bar{\nu}_{lL}\gamma^{m}\nu_{lL}
\left[ Z^{0}_{m}-\frac{1}{\sqrt{3}}\frac{1}{\sqrt{h(t)}}Z^{\prime 0}_{m} 
\right] ,
\label{e27}
\end{equation}
with 
\begin{equation}
h(t)=1+4t^{2},
\end{equation} 
for neutrinos and
\begin{equation}
{\cal L}_{l}^{NC} =-\frac{g}{4}\frac{M_{Z^{0}}}{M_{W^{\pm}}} 
\left[ \bar{l}\gamma^{m}(v_{l}+a_{l}\gamma^{5})lZ_{m}+ 
\bar{l}\gamma^{m}(v^{\prime}_{l}+a^{\prime}_{l}\gamma^5)lZ^{\prime 0}_{m} 
\right] ,
\label{e28}
\end{equation}
for the charged leptons, where we have defined
\begin{eqnarray}
\begin{array}{clccc}
v_{l}= & - \frac{1}{h(t)},&a_{l}=& 1,& \nonumber \\
v^{\prime}_{l}=& -\sqrt{ \frac{3}{h(t)}},
& a^{\prime}_{l}=& \frac{v^{\prime}_{l}}{3}.& \nonumber
\end{array}
\nonumber
\end{eqnarray}
We can use muon collider to discover the new neutral $Z^{\prime}$ boson using 
the reaction $\mu e \to \mu e$ it was show at 
\cite{Montero:1999en,Montero:1998sv} that $A_{RL}(\mu e)$ asymmetry is 
considerably enhanced, some phenomenological analyses was presented by 
references \cite{331susy1,mcr,Rodriguez:2010tn}.

For some phenomenological analyses, it is useful to define the following 
parameters \cite{Osland:2020onj,Osland:2022ryb}
\footnote{We want to perform similar constraints as presented in 
\cite{Osland:2020onj,Osland:2022ryb} in the MSUSY331.} 
\begin{equation}
\xi_{WW^{\prime}}=\left( \frac{M_{W}}{M_{W^{\prime}}} \right)^{2}, \,\ 
\xi_{ZZ^{\prime}}=\left( \frac{M_{Z}}{M_{Z^{\prime}}} \right)^{2}.
\label{Serenkova}
\end{equation}
We show some values of these parameters in MSUSY331 in our 
Tabs.(\ref{tabserenkovah1},\ref{tabserenkovah2}).

\begin{table}
\begin{tabular}{|c|c|c|}
\hline 
$v_{\chi}$ (GeV) & $\xi_{WW^{\prime}}$  & $\xi_{ZZ^{\prime}}$  \\
\hline 
1000 & $4.09 \times 10^{-4}$ & $2.03 \times 10^{-3}$ \\
1500 & $2.04 \times 10^{-4}$ & $1.25 \times 10^{-3}$ \\
2000 & $1.02 \times 10^{-4}$ & $8.14 \times 10^{-4}$ \\
2500 & $5.42 \times 10^{-5}$ & $5.61 \times 10^{-4}$ \\
3000 & $3.04 \times 10^{-5}$ & $4.07 \times 10^{-4}$ \\
3500 & $1.81 \times 10^{-5}$ & $3.07 \times 10^{-4}$ \\
4000 & $1.13 \times 10^{-5}$ & $2.39 \times 10^{-4}$ \\
4500 & $7.45 \times 10^{-6}$ & $1.91 \times 10^{-4}$ \\
5000 & $5.06 \times 10^{-6}$ & $1.56 \times 10^{-4}$ \\
\hline
\end{tabular}
\caption{The parameters $\xi_{WW^{\prime}}$ and $\xi_{ZZ^{\prime}}$, see 
Eq.(\ref{Serenkova}), as function of $v_{\chi}$ in GeV and we 
take $v_{\chi^{\prime}}=1000$ GeV and the others VEV are given in 
Eqs.(63,64) and Eq.(73) \cite{Rodriguez:2022hsj}.}
\label{tabserenkovah1}
\end{table}

\begin{table}
\begin{tabular}{|c|c|c|}
\hline 
$v_{\chi}$ (GeV) & $\xi_{WW^{\prime}}$  & $\xi_{ZZ^{\prime}}$  \\
\hline 
1000 & $4.09 \times 10^{-4}$ & $2.03 \times 10^{-3}$ \\
1500 & $2.04 \times 10^{-4}$ & $1.25 \times 10^{-3}$ \\
2000 & $1.02 \times 10^{-4}$ & $8.14 \times 10^{-4}$ \\
2500 & $5.42 \times 10^{-5}$ & $5.61 \times 10^{-4}$ \\
3000 & $3.04 \times 10^{-5}$ & $4.07 \times 10^{-4}$ \\
3500 & $1.81 \times 10^{-5}$ & $3.07 \times 10^{-4}$ \\
4000 & $1.13 \times 10^{-5}$ & $2.39 \times 10^{-4}$ \\
4500 & $7.45 \times 10^{-6}$ & $1.91 \times 10^{-4}$ \\
5000 & $5.06 \times 10^{-6}$ & $1.56 \times 10^{-4}$ \\
\hline
\end{tabular}
\caption{The parameters $\xi_{WW^{\prime}}$ and $\xi_{ZZ^{\prime}}$, see 
Eq.(\ref{Serenkova}), as function of $v_{\chi}$ in GeV and we 
take $v_{\chi^{\prime}}=1000$ GeV and the others VEV are given in 
Eqs.(63,64) and Eq.(73) \cite{Rodriguez:2022hsj}.}
\label{tabserenkovah2}
\end{table}

We would like to highlight, the new exotic quarks, $J$, $j_{1}$ 
and $j_{2}$, may be discovered by the Large Hadron Collider (LHC) thought 
$pp$ collisions, via the following subprocess\footnote{I want to thank Alexander S. Belyaev, who brought this process to my attention at the 
end of my PhD studies at IFT-Unesp, but unfortunately we were unable to publish this study together.}
\begin{eqnarray}
g+d \rightarrow U^{--}+J, \,\
g+u \rightarrow  U^{--}+j_{\alpha}, 
\label{intersting}
\end{eqnarray} 
its signature is $llXX$ and it can be detected at LHC if they really exist in 
nature \cite{Rodriguez:2010tn,dutta}. Another interesting process, with the same previous signature, to prove this model is
\begin{eqnarray}
e^{-}+e^{-} \rightarrow \tilde{\chi}^{--}+ \tilde{\chi}^{0}, 
\end{eqnarray} 
it was presented by\footnote{I want to thank M. Capdequi-Peyran\`ere, because he taught 
me a lot about the phenomenology of supersymmetric models in 2000 when I started 
studying this model.} \cite{331susy1,mcr,Rodriguez:2010tn}.

\section{Review Scalar Potential at m331}
\label{psusy1}

The most general scalar potential involving triplets and the 
anti-sextet is \cite{DeConto:2015eia}
\begin{eqnarray}
V(\eta,\rho,\chi,S)&=&V^{(2)}+V^{(3)}+ 
V^{(4a)}+V^{(4b)}+V^{(4c)}+V^{(4d)}+
V^{(4e)},
\end{eqnarray}
where
\begin{eqnarray}
V^{(2)}&=&
\mu_{1}^{2}(\eta^{\dagger}\eta) +
\mu_{2}^{2}(\rho^{\dagger}\rho) +
\mu_{3}^{2}(\chi^{\dagger}\chi) +
\mu_{4}^{2}Tr[(S^{\dagger} S)], \nonumber \\
V^{(3)}&=&
\frac{f_{1}}{3!}\varepsilon_{ijk}
\eta_{i}\rho_{j}\chi_{k}+
f_{2} \rho_{i}\chi_{j}S^{\dagger ij}+ 
f_{3}\eta_{i}\eta_{j}S^{\dagger ij}+
\frac{f_{4}}{3!}
\epsilon_{ijk}\epsilon_{lmn}
S_{il}S_{jm}S_{kn}+hc, \nonumber \\
V^{(4a)}&=&
a_{1}(\eta^{\dagger}\eta)^{2} +
a_{2}(\rho^{\dagger}\rho)^{2}+
a_{3}(\chi^{\dagger}\chi)^{2}+
(\chi^{\dagger}\chi)\left[
a_{4}(\eta^{\dagger}\eta) +
a_{5}(\rho^{\dagger}\rho ) \right] +
a_{6}(\eta^{\dagger}\eta)(\rho^{\dagger}\rho) \nonumber \\ 
&+&
a_{7}(\chi^{\dagger}\eta)
(\eta^{\dagger}\chi) +
a_{8}(\chi^{\dagger}\rho)
(\rho^{\dagger}\chi) +
a_{9}(\eta^{\dagger}\rho)
(\rho^{\dagger}\eta) +
\left[
a_{10}(\chi^{\dagger}\eta)
(\rho^{\dagger}\eta)+hc
\right], \nonumber \\
V^{(4b)}&=&
b_{1}(\chi^{\dagger}S)(\hat{\chi}\eta) +
b_{2}(\rho^{\dagger}S)(\hat{\rho}\eta) +
b_{3}(\eta^{\dagger}S) \left[
(\hat{\chi}\rho) - (\hat{\rho}\chi)
\right]+hc , \nonumber \\
V^{(4c)}&=&
c_{1}Tr[ (\hat{\eta}S) (\hat{\eta}S)] +
c_{2}Tr[ (\hat{\rho}S) (\hat{\rho}S)] + 
hc , \nonumber \\
V^{(4d)}&=&
d_{1}(\chi^{\dagger}\chi)Tr[(S^{\dagger}S)] +
d_{2} Tr[(\chi^{\dagger}S)(S^{\dagger}\chi)] +
d_{3}(\eta^{\dagger}\eta)Tr[(S^{\dagger}S)] +
d_{4} Tr[(\eta^{\dagger}S)(S^{\dagger}\eta)] 
\nonumber \\
&+&
d_{5}(\rho^{\dagger}\rho)Tr[(S^{\dagger}S)] +
d_{6} Tr[(\rho^{\dagger}S)(S^{\dagger}\rho)], \nonumber \\
V^{(4e)}&=&
e_{1}(Tr[(S^{\dagger}S)])^{2} +
e_{2}Tr[(S^{\dagger}S)(S^{\dagger}S)],
\label{potentialm331}
\end{eqnarray}
and we have defined the $V^{(4b)}$ and 
$V^{(4c)}$ terms
\begin{eqnarray}
\hat{x}_{ij}\equiv \epsilon_{ijk}x_{k}, 
\end{eqnarray}
with $x= \eta, \rho, \chi$. The constants 
$f_{i}$, $i=1,2,3$ and $4$ have dimension of mass.

\section{Construction Scalar Potential}
\label{constpot}

To get the scalar potential of our model we have to eliminate the auxiliarly 
fields $F$ and $D$ that appear in our model. We are going to pick up the $F$ 
and $D$- terms we get
\begin{eqnarray}
{\cal L}^{Gauge}_{D}&=&\frac{1}{2}D^{a}D^{a}+ \frac{1}{2}DD \,\ , \nonumber \\
{\cal L}^{Scalar}_{F}&=& \vert F_{\eta} \vert^2+ \vert F_{\rho} \vert^2+ 
\vert F_{\chi} \vert^2+ \vert F_{S} \vert^2 + 
\vert F_{\eta^{\prime}} \vert^2+ 
\vert F_{\rho^{\prime}} \vert^2+ 
\vert F_{\chi^{\prime}} \vert^2+ \vert F_{S^{\prime}} \vert^2, \nonumber \\
{\cal L}^{Scalar}_{D}&=& \frac{g}{2} \left[ \bar{\eta}\lambda^a\eta+ 
\bar{\rho}\lambda^a\rho+ \bar{\chi}\lambda^a\chi+ 
\bar{S}\lambda^aS- 
\bar{\eta}^{\prime}\lambda^{* a}\eta^{\prime}- 
\bar{\rho}^{\prime}\lambda^{* a}\rho^{\prime}- 
\bar{\chi}^{\prime}\lambda^{* a}\chi^{\prime}- 
\bar{S}^{\prime}\lambda^{* a}S^{\prime} \right] D^{a} \nonumber \\
&+& \frac{g^{ \prime}}{2} \left[  
\bar{\rho}\rho- \bar{\chi}\chi- \bar{\rho}^{\prime}\rho^{\prime}+ 
\bar{\chi}^{\prime}\chi^{\prime} \right]D, \nonumber \\
{\cal L}^{W2}_{F}&=&  
\mu_{ \eta} ( \eta F_{\eta^{\prime}}+ \eta^{\prime} F_{ \eta})+ 
\mu_{ \rho} ( \rho F_{\rho^{\prime}}+ \rho^{\prime} F_{ \rho})  +  
\mu_{ \chi} ( \chi F_{\chi^{\prime}}+ \chi^{\prime} F_{ \chi}) + hc, 
\nonumber \\
{\cal L}^{W3}_{F}&=& f_{1} \epsilon (F_{ \rho} \chi \eta+ \rho F_{ \chi} \eta+ \rho \chi F_{ \eta})  + 
f_{2}(2F_{\eta}\eta S+ \eta \eta F_{S}) +
f_{3}(F_{\rho}S \chi + \rho F_{S}\chi + \rho S F_{\chi}) \nonumber \\ &+&
3f_{4} \epsilon \epsilon SSF_{S}+f^{\prime}_{1} \epsilon (
F_{ \rho^{\prime}} \chi^{\prime} \eta^{\prime}+ 
\rho^{\prime} F_{ \chi^{\prime}} \eta^{\prime}+ 
\rho^{\prime} \chi^{\prime} F_{ \eta^{\prime}})
+f^{\prime}_{2}(2F_{\eta^{\prime}}\eta^{\prime} S^{\prime}+ 
\eta^{\prime} \eta^{\prime} F_{S^{\prime}}) \nonumber \\ &+&
f^{\prime}_{3}(F_{\rho^{\prime}}S^{\prime} \chi^{\prime} + 
\rho^{\prime} F_{S^{\prime}}\chi^{\prime} + 
\rho^{\prime} S^{\prime} F_{\chi^{\prime}})+
3f^{\prime}_{4} \epsilon \epsilon S^{\prime}S^{\prime}F_{S^{\prime}}+hc . 
\nonumber \\
\end{eqnarray}

From the equation described above we can construct
\begin{eqnarray}
{\cal L}_{F}&=&{\cal L}^{Scalar}_{F}+
{\cal L}^{W2}_{F}+{\cal L}^{W3}_{F} 
\nonumber \\
&=&\vert F_{\eta} \vert^2+ \vert F_{\rho} \vert^2+ 
\vert F_{\chi} \vert^2+  
\vert F_{\eta^{\prime}} \vert^2+ 
\vert F_{\rho^{\prime}} \vert^2+ 
\vert F_{\chi^{\prime}} \vert^2 \nonumber \\
&+&  \mu_{ \eta} ( \eta F_{\eta^{\prime}}+ \eta^{\prime} F_{ \eta})+ 
\mu_{ \rho} ( \rho F_{\rho^{\prime}}+ \rho^{\prime} F_{ \rho})  +  
\mu_{ \chi} ( \chi F_{\chi^{\prime}}+ \chi^{\prime} F_{ \chi}) 
\nonumber \\ &+&  
f_{1} \epsilon (F_{ \rho} \chi \eta+ \rho F_{ \chi} \eta+ \rho \chi F_{ \eta})  + 
f_{2}(2F_{\eta}\eta S+ \eta \eta F_{S}) +
f_{3}(F_{\rho}S \chi + \rho F_{S}\chi + \rho S F_{\chi}) \nonumber \\ &+&
3f_{4} \epsilon \epsilon SSF_{S}+f^{\prime}_{1} \epsilon (
F_{ \rho^{\prime}} \chi^{\prime} \eta^{\prime}+ 
\rho^{\prime} F_{ \chi^{\prime}} \eta^{\prime}+ 
\rho^{\prime} \chi^{\prime} F_{ \eta^{\prime}})
+f^{\prime}_{2}(2F_{\eta^{\prime}}\eta^{\prime} S^{\prime}+ 
\eta^{\prime} \eta^{\prime} F_{S^{\prime}}) \nonumber \\ &+&
f^{\prime}_{3}(F_{\rho^{\prime}}S^{\prime} \chi^{\prime} + 
\rho^{\prime} F_{S^{\prime}}\chi^{\prime} + 
\rho^{\prime} S^{\prime} F_{\chi^{\prime}})+
3f^{\prime}_{4} \epsilon \epsilon S^{\prime}S^{\prime}F_{S^{\prime}}+hc
\,\ , \nonumber \\
{\cal L}_{D}&=&{\cal L}^{Gauge}_{D}+
{\cal L}^{Scalar}_{D} \nonumber \\
&=& \frac{1}{2}D^{a}D^{a}+ \frac{1}{2}DD 
+ \frac{g}{2} \left[ \bar{\eta}\lambda^a\eta+ 
\bar{\rho}\lambda^a\rho+ \bar{\chi}\lambda^a\chi- 
\bar{\eta}^{\prime}\lambda^{* a}\eta^{\prime}- 
\bar{\rho}^{\prime}\lambda^{* a}\rho^{\prime} \right. \nonumber \\
&-& \left. 
\bar{\chi}^{\prime}\lambda^{* a}\chi^{\prime} \right] D^{a}+ 
\frac{g^{ \prime}}{2} \left[  
\bar{\rho}\rho- \bar{\chi}\chi- \bar{\rho}^{\prime}\rho^{\prime}+ 
\bar{\chi}^{\prime}\chi^{\prime} \right]D.
\label{auxiliarm1}
\end{eqnarray}

We will now show that these fields can be eliminated through the 
Euler-Lagrange equations
\begin{eqnarray}
\frac{\partial {\cal L}}{\partial \phi}- \partial_{m} 
\frac{\partial {\cal L}}{\partial (\partial_{m} \phi)}=0 \,\ ,  
\label{Euler-Lagrange Equation}
\end{eqnarray}
where 
$\phi = \eta , \rho , \chi ,S, \eta^{\prime}, \rho^{\prime}, \chi^{\prime}, 
S^{\prime}$. 
Formally auxiliary fields are defined 
as fields having no kintetic terms. Thus, this definition immediately yields 
that the Euler-Lagrange equations for auxiliary fields simplify to 
$\frac{\partial {\cal L}}{\partial \phi}=0$.

\begin{eqnarray}
V_D&=&-{\cal L}_D=\frac{1}{2}\left(D^aD^a+DD\right)\nonumber \\ &=&
\frac{g^{\prime2}}{2}(\rho^\dagger\rho-\rho^{\prime\dagger}\rho^\prime
-\chi^\dagger\chi+\chi^{\prime\dagger}\chi^\prime)^2+
\frac{g^2}{8}\sum_{i,j}\left(\eta^\dagger_i\lambda^a_{ij}\eta_j
+\rho^\dagger_i\lambda^a_{ij}\rho_j
+\chi^\dagger_i\lambda^a_{ij}\chi_j+S^\dagger_{ij}\lambda^a_{jk}
S_{kl}\right.
\nonumber \\ &-&
\left.\eta^{\prime\dagger}_i\lambda^{*a}_{ij}\eta^\prime_j 
-\rho^{\prime\dagger}_i\lambda^{*a}_{ij}\rho^\prime_j
-\chi^{\prime\dagger}_i\lambda^{*a}_{ij}\chi^\prime_j-
S^{\prime\dagger}_{ij}\lambda^{*a}_{jk} S^\prime_{kl} \right)^2,
\label{esd}
\end{eqnarray}
We can use the following relations
\begin{eqnarray}
\lambda^{a}_{ij}\lambda^{a}_{kl}= 
\frac{(-2)}{3}\delta_{ij}\delta_{kl}+
2\delta{il}\delta_{jk}.
\end{eqnarray}
Then
\begin{eqnarray}
\eta^{\dagger}_{i}\eta_{j}
\eta^{\dagger}_{k}\eta_{l}
\lambda^{a}_{ij}\lambda^{a}_{kl}=- 
\frac{2}{3}
\left( \eta^{\dagger}\eta \right)
\left( \eta^{\dagger}\eta \right)+ 2 
\left( \eta^{\dagger}\eta \right)
\left( \eta^{\dagger}\eta \right)= 
\frac{4}{3} \left( \eta^{\dagger}\eta \right)^{2},
\end{eqnarray}
in similar way
\begin{eqnarray}
\rho^{\dagger}_{i}\rho_{j}
\rho^{\dagger}_{k}\rho_{l}
\lambda^{a}_{ij}\lambda^{a}_{kl}&=& 
\frac{4}{3} \left( \rho^{\dagger}\rho \right)^{2}, \nonumber \\
\chi^{\dagger}_{i}\chi_{j}
\chi^{\dagger}_{k}\chi_{l}
\lambda^{a}_{ij}\lambda^{a}_{kl}&=& 
\frac{4}{3} \left( \chi^{\dagger}\chi \right)^{2}, \nonumber \\
S^{\dagger}_{il}S_{mj}
S^{\dagger}_{kn}S_{op}
\lambda^{a}_{lm}\lambda^{a}_{no}&=&- 
\frac{2}{3}
{\mbox Tr}\left[ 
\left( S^{\dagger}S \right)
\left( S^{\dagger}S \right) \right] + 2 
{\mbox Tr} \left[ (S^{\dagger}S) 
(S^{\dagger}S) \right].
\end{eqnarray}

By another hand
\begin{eqnarray}
\eta^{\dagger}_{i}\eta_{j}
\rho^{\dagger}_{k}\rho_{l}
\lambda^{a}_{ij}\lambda^{a}_{kl}&=&- 
\frac{2}{3}
\left( \eta^{\dagger}\eta \right)
\left( \rho^{\dagger}\rho \right)+ 2 
\left( \rho^{\dagger}\eta \right)
\left( \eta^{\dagger}\rho \right), 
\nonumber \\
\eta^{\dagger}_{i}\eta_{j}
\chi^{\dagger}_{k}\chi_{l}
\lambda^{a}_{ij}\lambda^{a}_{kl}&=&- 
\frac{2}{3}
\left( \eta^{\dagger}\eta \right)
\left( \chi^{\dagger}\chi \right)+ 2 
\left( \chi^{\dagger}\eta \right)
\left( \eta^{\dagger}\chi \right), 
\nonumber \\
\rho^{\dagger}_{i}\rho_{j}
\chi^{\dagger}_{k}\chi_{l}
\lambda^{a}_{ij}\lambda^{a}_{kl}&=&- 
\frac{2}{3}
\left( \rho^{\dagger}\rho \right)
\left( \chi^{\dagger}\chi \right)+ 2 
\left( \chi^{\dagger}\rho \right)
\left( \rho^{\dagger}\chi \right),
\end{eqnarray}

We can rewrite $V_{D}$ in the following way
\begin{eqnarray}
V_{D}&=&\frac{g^{\prime2}}{2}
\left[ (\rho^{\dagger}\rho)^{2}+
(\chi^{\dagger}\chi)^{2}-
2(\rho^{\dagger}\rho)(\chi^{\dagger}\chi)
+ \ldots  \right] \nonumber \\
&+& \frac{g^{2}}{8}\left\{
\frac{(-4)}{3}\left[
(\eta^{\dagger}\eta)^{2}+
(\rho^{\dagger}\rho)^{2}+
(\chi^{\dagger}\chi)^{2}+
(\eta^{\dagger}\eta)(\rho^{\dagger}\rho)+
(\eta^{\dagger}\eta)(\chi^{\dagger}\chi)
\right. \right. \nonumber \\ 
&+& \left. \left.
(\eta^{\dagger}\eta)Tr[(S^{\dagger}S)]+
(\rho^{\dagger}\rho)(\chi^{\dagger}\chi)
+
(\rho^{\dagger}\rho)Tr[(S^{\dagger}S)]+ 
(\chi^{\dagger}\chi)Tr[(S^{\dagger}S)]
\right]
\right. \nonumber \\
&+& \left. 
4 \left[
(\eta^{\dagger}\rho)(\rho^{\dagger}\eta)+
(\eta^{\dagger}\chi)(\chi^{\dagger}\eta)
+
Tr[(S^{\dagger}\eta)(\eta^{\dagger}S)]+
(\rho^{\dagger}\chi)(\chi^{\dagger}\rho)
\right. \right. \nonumber \\
&+& \left. \left.
Tr[(S^{\dagger}\rho)(\rho^{\dagger}S)]+
Tr[(S^{\dagger}\chi)(\chi^{\dagger}S)]
\right] 
- \frac{2}{3}
\left(Tr[(S^{\dagger}S)]\right)^{2}
\right.  \nonumber \\
&+& \left.  2 
Tr[(S^{\dagger}S)(S^{\dagger}S)]
+ \ldots
\right\}.
\end{eqnarray}
where $\ldots$ include the new fields 
due the SUSY algebra. As we want to 
compare our potential with the m331 
we omitt those terms.

\begin{eqnarray}
V_F&=&-{\cal L}_F=\sum_mF^*_m F_m\nonumber \\ &=&
\sum_{i,j,k}\left[\left\vert \mu_\eta \eta^{\prime}_i+f_1
\epsilon_{ijk}\rho_j\chi_k
+2f_2 \eta_iS_{ij}\right\vert^2+
\left\vert \mu_\rho \rho^\prime_i+f_1
\epsilon_{ijk}\chi_j\eta_k+f_3 \chi_i S_{ij}\right\vert^2
\right.\nonumber \\ &+&\left.
\left\vert
\mu_\chi \chi^\prime_i+f_1 \epsilon_{ijk}\rho_j\eta_k 
+f_3 \rho_i S_{ij}\right\vert^2
+\left\vert \mu_S S^\prime_{ij}+f_2 \eta_i\eta_j+f_3
\chi_i\rho_j +3f_{4}\epsilon_{ilk} \epsilon_{jmn}S_{lm}S_{kn} 
\right\vert^2\right.\nonumber \\ &+&\left.\left\vert 
\mu_\eta \eta_i+f^\prime_1
\epsilon_{ijk}\rho^\prime_j\chi^\prime_k
+2f^\prime_2\eta^\prime_iS^\prime_{ij}\right\vert^2+
\left\vert \mu_\rho \rho_i+f^\prime_1
\epsilon_{ijk}\chi^\prime_j\eta^\prime_k+f^\prime_3\chi^\prime_i
S^\prime_{ij} 
\right\vert^2\right.\nonumber \\ &+&\left.
\left\vert
\mu_\chi\chi_i+f^\prime_1 \epsilon_{ijk}\rho^\prime_j
\eta^\prime_k
+f^\prime_3 \rho^\prime_i S^\prime_{ij}\right\vert^2
+\left\vert
\mu_S S_{ij}+f^\prime_2 \eta^\prime_i\eta^\prime_j+ 
f^\prime_3
\chi^\prime_i\rho^\prime_j 
+3f^{\prime}_{4}\epsilon_{ilk} \epsilon_{jmn}S^{\prime}_{lm}S^{\prime}_{kn}
\right\vert^2
\right] \nonumber \\
\end{eqnarray}
We can use the following relation
\begin{equation}
\epsilon_{ijk}\epsilon_{kmn}=
\delta_{im}\delta_{jn}-
\delta_{in}\delta_{jm}.
\end{equation}

As we want to get the M331 we will make\footnote{To satisfy ${\cal F}$ 
defined in Eq.(\ref{fquantumnumber}).} 
\begin{equation}
f_{2}=f_{4}=0=f^{\prime}_{2}=f^{\prime}_{4},
\end{equation} 
then we get
\begin{eqnarray}
\left\vert 
\mu_\eta \eta_i + \ldots 
\right\vert^2 
&=& \mu_{\eta}
(\eta^{\dagger}\eta), \,\
\left\vert f_1
\epsilon_{ijk}\rho_j\chi_k + 
\ldots \right\vert^2 =
f^{2}_{1}\left[
(\rho^{\dagger}\rho)(\chi^{\dagger}\chi)- 
(\rho^{\dagger}\chi)(\chi^{\dagger}\rho) 
\right], \nonumber \\
\left\vert 
\mu_\rho \rho_i + \ldots 
\right\vert^2 
&=& \mu^{2}_{\rho}
(\rho^{\dagger}\rho), \,\ 
\left\vert 
\mu_\chi \chi_i + \ldots 
\right\vert^2 
= \mu^{2}_{\chi}
(\chi^{\dagger}\chi), \nonumber \\
\left\vert f_1 \epsilon_{ijk}\chi_j\eta_k+
f_3 \chi_i S_{ij}+ \ldots \right\vert^2 
&=& f^{2}_{1} \left[ 
( \chi^{\dagger}\chi )
( \eta^{\dagger}\eta )-
( \chi^{\dagger}\eta )
( \eta^{\dagger}\chi ) 
\right]+
f_{1}f_{3} \left[ 
\epsilon ( \chi^{\dagger}\eta^{\dagger})
( \chi S)+hc 
\right] \nonumber \\ &+&
f^{2}_{3} {\mbox Tr}\left[ 
(S^{\dagger}\chi )( \chi^{\dagger}S) 
\right], \nonumber \\
\left\vert f_1 \epsilon_{ijk}\rho_j\eta_k+
f_3 \rho_i S_{ij}+ \ldots \right\vert^2 
&=& f^{2}_{1} \left[ 
( \rho^{\dagger}\rho )
( \eta^{\dagger}\eta )-
( \rho^{\dagger}\eta )
( \eta^{\dagger}\rho ) 
\right]+
f_{1}f_{3} \left[ 
\epsilon ( \rho^{\dagger}\eta^{\dagger})
( \rho S)+hc 
\right] \nonumber \\ &+&
f^{2}_{3} {\mbox Tr}\left[ 
(S^{\dagger}\rho )( \rho^{\dagger}S) 
\right], \nonumber \\
\left\vert
\mu_S S_{ij}+ \ldots 
\right\vert^2 &=& \mu^{2}_S 
(Tr[(S^{\dagger}S)])^{2}, \nonumber \\ 
\left\vert f_1
\chi_i\rho_j + \ldots \right\vert^2 
&=& f^{2}_1 \left[
(\rho^{\dagger}\rho)(\chi^{\dagger}\chi)-
(\rho^{\dagger}\chi)(\chi^{\dagger}\rho)
\right]. \nonumber \\
\end{eqnarray}

Therefore
\begin{eqnarray}
V_F&=& 
\mu^{2}_{\eta}(\eta^{\dagger}\eta)^{2}+
\mu^{2}_{\rho}(\rho^{\dagger}\rho)^{2}+
\mu^{2}_{\chi}(\chi^{\dagger}\chi)^{2}+ 
\mu^{2}_{S} (Tr[(S^{\dagger}S)])^{2}
\nonumber \\ &+&
f^{2}_{1}\left[
(\eta^{\dagger}\eta)(\rho^{\dagger}\rho)-
(\rho^{\dagger}\eta)(\eta^{\dagger}\rho)+
(\eta^{\dagger}\eta)(\chi^{\dagger}\chi)
\right. \nonumber \\
&-& \left.
(\chi^{\dagger}\eta)(\eta^{\dagger}\chi)+
(\rho^{\dagger}\rho)(\chi^{\dagger}\chi)-
(\rho^{\dagger}\chi)(\chi^{\dagger}\rho)
\right]+ 
f^{2}_{3}\left\{
Tr[(S^{\dagger}\rho)(\rho^{\dagger}S)]+
Tr[(S^{\dagger}\chi)(\chi^{\dagger}S)]
\right\}
\nonumber \\ &+&
\left\{ f_{1}f_{3}
\left[ 
\epsilon (\rho^{\dagger}\eta^{\dagger}) 
(\rho S)+
\epsilon (\chi^{\dagger}\eta^{\dagger}) 
(\chi S) \right]+hc \right\}
+ \ldots ,
\label{esf}
\end{eqnarray}

The scalar potential in this model is
\begin{eqnarray}
V_{scalar}&=&
m^{2}_{ \eta}(\eta^{ \dagger}\eta)+
m^{2}_{ \rho}(\rho^{ \dagger}\rho)+
m^{2}_{ \chi}(\chi^{ \dagger}\chi)+
m^{2}_{S}Tr[(S^{ \dagger}S)]
-[k_{1}\epsilon \rho \chi \eta +
k_{3}\chi \rho S^{\dagger}+ hc] \nonumber \\
&+&
\left( \mu^{2}_{\eta}- \frac{g^{2}}{6}
\right) (\eta^{\dagger}\eta)^{2}+
\left( \mu^{2}_{\rho}+ 
\frac{g^{\prime 2}}{2}- \frac{g^{2}}{6}
\right) (\rho^{\dagger}\rho)^{2}+
\left( \mu^{2}_{\chi}+ 
\frac{g^{\prime 2}}{2}- \frac{g^{2}}{6}
\right) (\chi^{\dagger}\chi)^{2}
\nonumber \\ &+&
\left(
f^{2}_{1}- \frac{g^{2}}{6}
\right)(\eta^{\dagger}\eta)(\rho^{\dagger}\rho)+
\left(
f^{2}_{1}- \frac{g^{2}}{6}
\right)(\eta^{\dagger}\eta)(\chi^{\dagger}\chi)+
\left(
f^{2}_{1}-g^{\prime 2}- \frac{g^{2}}{6}
\right)(\rho^{\dagger}\rho)(\chi^{\dagger}\chi)
\nonumber \\ &+&
\left(
\frac{g^{2}}{2}-f^{2}_{1}
\right)(\eta^{\dagger}\rho)(\rho^{\dagger}\eta)+
\left(
\frac{g^{2}}{2}-f^{2}_{1}
\right)(\eta^{\dagger}\chi)(\chi^{\dagger}\eta)+
\left(
\frac{g^{2}}{2}-f^{2}_{1}
\right)(\rho^{\dagger}\chi)(\chi^{\dagger}\rho) \nonumber \\
&-& \frac{g^{2}}{6} \left[
(\eta^{\dagger}\eta)Tr[(S^{\dagger}S)]+
(\rho^{\dagger}\rho)Tr[(S^{\dagger}S)]+ 
(\chi^{\dagger}\chi)Tr[(S^{\dagger}S)] 
\right] \nonumber \\
&+& 
\frac{g^{2}}{2}
Tr[(S^{\dagger}\eta)(\eta^{\dagger}S)]+
\left( \frac{g^{2}}{2}+
f^{2}_{3} \right)
\left\{
Tr[(S^{\dagger}\rho)(\rho^{\dagger}S)]+
Tr[(S^{\dagger}\chi)(\chi^{\dagger}S)]
\right\} \nonumber \\
&+&
\left( \mu^{2}_{S}-
\frac{g^{2}}{2} \right)
\left(Tr[(S^{\dagger}S)]\right)^{2}
+g^{2} Tr[(S^{\dagger}S)(S^{\dagger}S)] 
\nonumber \\
&+&
\left\{ f_{1}f_{3}
\left[ 
\epsilon (\rho^{\dagger}\eta^{\dagger}) 
(\rho S)+
\epsilon (\chi^{\dagger}\eta^{\dagger}) 
(\chi S) \right]+H.c.\right\}
+ \ldots
\end{eqnarray}
and comparing it with Eq.(\ref{potentialm331}) we can see
\begin{eqnarray}
\mu^{2}_{1}&=& m^{2}_{ \eta}, \quad 
\mu^{2}_{2}= m^{2}_{ \rho}, \quad
\mu^{2}_{3}= m^{2}_{ \chi}, \quad
\mu^{2}_{4}= m^{2}_{S}, \nonumber \\ 
f_{1}&=&-k_{1}, \quad
f_{2}=-k_{3},
\nonumber \\
a_{1}&=&\mu^{2}_{\eta}- \frac{g^{2}}{6}, 
\quad 
e_{1}=\mu^{2}_{S}-\frac{g^{2}}{2}, \quad
e_{2}=g^{2}, \nonumber \\
a_{2}&=&\mu^{2}_{\rho}+ \frac{g^{\prime 2}}{2}
- \frac{g^{2}}{6},
a_{3}=\mu^{2}_{\chi}+ \frac{g^{\prime 2}}{2}
- \frac{g^{2}}{6}, \nonumber \\
a_{4}&=&- \frac{g^{2}}{6}+
\frac{f^{2}_{1}}{9}=a_{6}=a_{7}=a_{8}=a_{9}, \nonumber \\
a_{5}&=&- g^{\prime 2}- \frac{g^{2}}{6}+
\frac{f^{2}_{1}}{9}, \nonumber \\
d_{1}&=&- \frac{g^{2}}{6}=d_{3}=d_{4}=d_{5}, \nonumber \\
d_{2}&=&d_{4}=d_{6}=\frac{g^{2}}{2}, 
\nonumber \\
a_{10}&=&=0.
\label{comppotnonsusycomsusy}
\end{eqnarray}


\end{document}